\documentstyle[12pt,a4,epsfig]{article}
\newcommand{\bq}{\begin{equation}}
\newcommand{\eq}{\end{equation}}
\newcommand{\beq}  {\begin{eqnarray}}
\newcommand{\eeq}  {\end{eqnarray}}
\newcommand{\rG}   {{\rm GUT}}
\newcommand{\MG}   {{\ifmmode M_\rG         \else $M_\rG$          \fi}}

\newcommand{\mh}{\mbox{$m_{\mathrm{h}}$}}
\newcommand{\mA}{\mbox{$m_{\mathrm{A}}$}}

\newcommand{\mb}   {{\ifmmode m_{b}         \else $m_{b}$          \fi}}
\newcommand{\mt}   {{\ifmmode m_{t}         \else $m_{t}$          \fi}}
\newcommand{\agut} {{\ifmmode \alpha_\rG    \else $\alpha_\rG$     \fi}}
\newcommand{\mgut} {{\ifmmode m_\rG         \else $m_\rG$          \fi}}
\newcommand{\mze}  {{\ifmmode m_0           \else $m_0$            \fi}}
\newcommand{\mha}  {{\ifmmode m_{1/2}       \else $m_{1/2}$        \fi}}
\newcommand{\tb}   {{\ifmmode \tan\beta     \else $\tan\beta$      \fi}}
\newcommand{\tanb}   {{\ifmmode \tan\beta     \else $\tan\beta$      \fi}}
\newcommand{\mz}   {{\ifmmode m_{Z}         \else $m_{Z}$          \fi}}
\newcommand{\ai}   {{\ifmmode \alpha_i      \else $\alpha_i$       \fi}}
\newcommand{\aii}  {{\ifmmode \alpha_i^{-1} \else $\alpha_i^{-1}$  \fi}}
\newcommand{\MSb}  {{\ifmmode \overline{\rm MS} \else
                             $\overline{\rm MS}$                   \fi}}
\newcommand{\DRb}  {{\ifmmode \overline{\rm DR} \else
      $\overline{\rm DR}$                   \fi}}
\newcommand{\DRbar}{{\ifmmode \overline{DR} \else $ \overline{DR}$ \fi}}
\newcommand{\msusy}{{\ifmmode M_{SUSY}      \else $M_{SUSY}$       \fi}}
\newcommand{\tal}  {{\ifmmode \tilde{\alpha} \else $\tilde{\alpha}$ \fi}}
\newcommand{\rb}[1]{\raisebox{1.5ex}[-1.5ex]{#1}}

\newcommand{\sws}  {{\ifmmode \;\sin^2\theta_W
                     \else    $\;\sin^{2}\theta_{W}$               \fi}}
\newcommand{\cws}  {{\ifmmode \;\cos^2\theta_W  
                     \else    $\;\cos^{2}\theta_{W}$               \fi}}
\newcommand{\sw}   {{\ifmmode\;\sin\theta_W\else $\sin\theta_{W}$  \fi}}
\newcommand{\cw}   {{\ifmmode\;\cos\theta_W\else $\;\cos\theta_{W}$\fi}}
\newcommand{\tw}   {{\ifmmode\;\tan\theta_W\else $\;\tan\theta_{W}$\fi}}
\newcommand{\bsg}  {{\ifmmode b\rightarrow s\gamma
                     \else $b\rightarrow s\gamma$ \fi}}
\newcommand{\Bbsg}  {{\ifmmode {\cal{BR}}(\b\rightarrow s\gamma)
                     \else ${\cal{BR}}(b\rightarrow s\gamma)$ \fi}}

\newcommand{\rPL}  {{\rm Planck}}
\newcommand{\mplanck} {{\ifmmode M_\rPL         \else $M_\rPL$          \fi}}
\newcommand{\rST}  {{\rm SO(10)}}
\newcommand{\msoten} {{\ifmmode M_\rST         \else $M_\rST$          \fi}}
\newcommand{\rpv}{{\ifmmode\;\not \!\! R_p\else $\;\not \!\! R_p$\fi}}
\newcommand{\etmiss}{{\ifmmode\;\not \!\! E_t\else $\;\not \!\! E_t$\fi}}

\def\be{\begin{equation}}
\def\ee{\end{equation}}
\def\bea{\begin{eqnarray}}
\def\eea{\end{eqnarray}}

\def\rG{{\rm GUT}}

\def\MG{M_\rG}

\newcommand{\ba}   {\begin{array}}
\newcommand{\ea}   {\end{array}}

\newcommand{\lnf}  {{\ifmmode \Lambda^{(N_f)} \else $\Lambda^{(N_f)}$\fi}}
\newcommand{\ms}   {{\ifmmode \overline{MS} \else $\overline{MS}$\fi}}
\newcommand{\dr}   {{\ifmmode \overline{DR} \else $\overline{DR}$\fi}}
\newcommand{\lms}  {{\ifmmode \Lambda^{(5)}_{\overline{MS}}
                            \else $\Lambda^{(5)}_{\overline{MS}}$\fi}}
\newcommand{\lam}  {{\ifmmode \Lambda \else $\Lambda$\fi}}
\newcommand{\gev}  {{\ifmmode {\rm GeV} \else ${\rm GeV}$\fi}}
\newcommand{\gevc} {{\ifmmode {\rm GeV/c^2} \else ${\rm GeV/c^2}$\fi}}
\newcommand{\tev}  {{\ifmmode {\rm TeV} \else ${\rm TeV}$\fi}}
\newcommand{\tevc} {{\ifmmode {\rm TeV/c^2} \else ${\rm TeV/c^2}$\fi}}
\newcommand{\lp}   {{\ifmmode L^+  \else $L^+$\fi}}
\newcommand{\lm}   {{\ifmmode L^-  \else $L^-$\fi}}
\newcommand{\mlp}  {{\ifmmode M(L^-)  \else $M(L^-)$\fi}}
\newcommand{\mlz}  {{\ifmmode M(L^0)  \else $M(L^0)$\fi}}
\newcommand{\lz}   {{\ifmmode L^0     \else $L^0$\fi}}
\newcommand{\ev}   {{\ifmmode GeV/c^2       else $GeV/c^2$\fi}}
\newcommand{\tri}  {{\ifmmode \triangleup  \else $\triangleup$\fi}}
\newcommand{\unl}  {{\ifmmode U_{lL^0}  \else $U_{lL^0}$\fi}}
\newcommand{\gL}   {{\ifmmode g_L  \else $g_{L}$\fi}}
\newcommand{\gR}   {{\ifmmode g_R  \else $g_{R}$\fi}}
\newcommand{\gumu} {{\ifmmode \gamma^{\mu}  \else $\gamma^{\mu}$\fi}}
\newcommand{\gunu} {{\ifmmode \gamma^{\nu}  \else $\gamma^{\nu}$\fi}}
\newcommand{\gdmu} {{\ifmmode \gamma_{\mu}  \else $\gamma_{\mu}$\fi}}
\newcommand{\gdnu} {{\ifmmode \gamma_{\nu}  \else $\gamma_{\nu}$\fi}}
\newcommand{\stw}  {{\ifmmode\sin^2\theta_W  \else $\sin^{2}\theta_{W}$\fi}}
\newcommand{\qq}   {{\ifmmode q\overline{q} \else $q\overline{q}$\fi}}
\newcommand{\lR}   {{\ifmmode l_R  \else $l_R$\fi}}
\newcommand{\lL}   {{\ifmmode l_L  \else $l_L$\fi}}
\newcommand{\nt}   {{\ifmmode \nu_{\tau} \else $\nu_{\tau}$\fi}}
\newcommand{\nuR}  {{\ifmmode \nu_R  \else $\nu_R$\fi}}
\newcommand{\nuL}  {{\ifmmode \nu_L  \else $\nu_L$\fi}}
\newcommand{\qR}   {{\ifmmode g_R  \else $q_R$\fi}}
\newcommand{\qL}   {{\ifmmode q_L  \else $q_L$\fi}}
\newcommand{\qRp}  {{\ifmmode q_R'  \else $q_{R}$'\fi}}
\newcommand{\qLp}  {{\ifmmode q_L'  \else $q_{L}$'\fi}}
\newcommand{\est}{{\ifmmode e^{\bf \ast}  \else $e^{\bf \ast}$\fi}}
\newcommand{\lst}{{\ifmmode l^{\bf \ast}  \else $l^{\bf \ast}$\fi}}
\newcommand{\must}{{\ifmmode \mu^{\bf \ast}  \else $\mu^{\bf \ast}$\fi}}
\newcommand{\taust}{{\ifmmode \tau^{\bf \ast}  \else $\tau^{\bf \ast}$
\fi}}
\newcommand{\pperp}{{\ifmmode p_t  \else $p_t$\fi}}
\newcommand{\et}{{\ifmmode E_t  \else $E_t$\fi}}
\newcommand{\xt}{{\ifmmode x_t  \else $x_t$\fi}}
\newcommand{\smumu}{{\ifmmode \sigma_{\mu\mu}  \else $\sigma_{\mu\mu}$
\fi}}
\newcommand{\eg}{{\ifmmode e\gamma  \else $e\gamma$\fi}}
\newcommand{\epem}{{\ifmmode e^+e^-  \else $e^+e^-$\fi}}
\newcommand{\lplm}{{\ifmmode L^+L^-  \else $L^+L^-$\fi}}
\newcommand{\pp}{{\ifmmode p\overline p  \else $p\overline p$\fi}}
\newcommand{\llz}{{\ifmmode L^0\overline{L}^0 \else
$L^0\overline{L}^0$\fi}}
\newcommand{\epemt}{{\ifmmode e^+e^- \to  \else $e^+e^- \to$\fi}}
\newcommand{\eb}{{\ifmmode E_{beam}  \else $E_{beam}$\fi}}
\newcommand{\ip}{{\ifmmode pb^{-1}  \else $pb^{-1}$\fi}}
\newcommand{\upm}{{\ifmmode ^{\pm}  \else $^{\pm}$\fi}}
\newcommand{\de}{{\ifmmode ^{\circ}  \else $^{\circ}$ \fi}}
\newcommand{\appr}{{\ifmmode \sim \else $\sim$ \fi}}
\newcommand{\corresp}{{\ifmmode \stackrel{\wedge}{=}
                      \else   $\stackrel{\wedge}{=}$ \fi}}
\newcommand{\sqrts}{{\ifmmode \sqrt{s} \else $\sqrt{s}$\fi}}
\newcommand{\zz}{{\ifmmode Z^0  \else $Z^0$\fi}}
\newcommand{\mzs}{{\ifmmode M_{Z}^2  \else $M_{Z}^2$\fi}}
\newcommand{\mws}{{\ifmmode M_{W}^2  \else $M_{W}^2$\fi}}
\newcommand{\gt}{{\ifmmode \Gamma_{tot} \else $\Gamma_{tot}$\fi}}
\newcommand{\msusys}{{\ifmmode M_{SUSY}^2  \else $M_{SUSY}^2$\fi}}
\newcommand{\su}{{\ifmmode SU(3)_C\otimes\- SU(2)_L\otimes\- U(1)_Y  \else $SU(3)_C\otimes SU(2)_L\otimes U(1)_Y$\fi}}
\newcommand{\suthree}{{\ifmmode SU(3)_C  \else $SU(3)_C$\fi}}
\newcommand{\sutwo}{{\ifmmode  SU(2)_L\otimes U(1)_Y \else $SU(2)_L\otimes U(1)_Y$\fi}}
\newcommand{\taup} {{\ifmmode \tau_{proton} \else $\tau_{proton}$\fi}}
\newcommand{\mguts}{{\ifmmode M_{GUT}^2  \else $M_{GUT}^2$\fi}}
\newcommand{\mts} {{\ifmmode m_{t}^2    \else $m_{t}^2$\fi}}

\newcommand{\mtau}{{\ifmmode m_{\tau}  \else $m_{\tau}$\fi}}
\newcommand{\dpp}{{\ifmmode \delta_{pert} \else $\delta_{pert}$\fi}}
\newcommand{\dnp}{{\ifmmode\delta_{non-pert}\else$\delta_{non-pert}$\fi}}
\newcommand{\dew}{{\ifmmode \delta_{\rm EW}\else $\delta_{\rm EW}$\fi}}
\newcommand{\rt}{{\ifmmode R_{\tau}  \else
                 $R_{\tau} $\fi}}
\newcommand{\rz}{{\ifmmode R_{Z}  \else
                 $R_{Z} $\fi}}

\newcommand{\swb}{{\ifmmode \sin^2\theta_{\overline{MS}}
                     \else $\sin^2\theta_{\overline{MS}}$\fi}}
\newcommand{\cwb}{{\ifmmode \cos^2\theta_{\overline{MS}}
                     \else $\cos^2\theta_{\overline{MS}}$\fi}}
\def\ai{\alpha_i}
\def\aii{\alpha_i^{-1}}

\def\rG{{\rm GUT}}
\def\rt{{\rm threshold}}

\def\MG{M_\rG}

\def\DRbar{{\overline{DR}}}

\newcommand{\LL}{{\ifmmode {\cal L} \else ${\cal L}$\fi}}
\newcommand{\hz}{{\ifmmode {\rm Hz} \else ${\rm Hz}$\fi}}
\newcommand{\khz}{{\ifmmode {\rm kHz} \else ${\rm kHz}$\fi}}
\newcommand{\mhz}{{\ifmmode {\rm mHz} \else ${\rm mHz}$\fi}}
\newcommand{\as}{{\ifmmode \alpha_s  \else $\alpha_s$\fi}}
\newcommand{\asmz}{{\ifmmode \alpha_s(M_Z) \else $\alpha_s(M_Z)$\fi}}
\newcommand{\astau}{{\ifmmode \alpha_s(M_{\tau})
                       \else $\alpha_s(M_{\tau})$\fi}}
\newcommand{\ca}{{\ifmmode C_a  \else $C_a$\fi}}
\newcommand{\tf}{{\ifmmode T_{\mbox{\scriptsize Fermion}}
           \else $T_{\mbox{\scriptsize Fermion}}$\fi}}
\newcommand{\ts}{{\ifmmode T_{\mbox{\scriptsize Scalar}}
           \else $T_{\mbox{\scriptsize Scalar}}$ \fi}}
\newcommand{\mhiggs}{{\ifmmode M_{\mbox{\scriptsize Higgs}}
           \else $M_{\mbox{\scriptsize Higgs}}$\fi}}
\newcommand{\mthres}{{\ifmmode M_{\mbox{\scriptsize threshold}}
           \else $M_{\mbox{\scriptsize threshold}}$ \fi}}
\newcommand{\msbar}{{\ifmmode \overline{MS} \else $\overline{MS}$\fi}}
\newcommand{\drbar}{{\ifmmode \overline{DR} \else $\overline{DR}$\fi}}
\newcommand{\lamms}{{\ifmmode \Lambda_{\overline{MS}}
                       \else $\Lambda_{\overline{MS}}$\fi}}
\newcommand{\rr}{{{\ifmmode {\cal R}_2 }\else ${\cal R}_2 $\fi}}
\newcommand{\rrr}{{{\ifmmode {\cal R}_3 }\else ${\cal R}_3 $\fi}}
\newcommand{\rrrr}{{{\ifmmode {\cal R}_4 }\else ${\cal R}_4 $\fi}}
\newcommand{\jdd}{{{\ifmmode {\cal D}_2 }\else ${\cal D}_2 $\fi}}
\newcommand{\jddd}{{{\ifmmode {\cal D}_3 }\else ${\cal D}_3 $\fi}}
\newcommand{\jdddd}{{{\ifmmode {\cal D}_4 }\else ${\cal D}_4 $\fi}}
\newcommand{\rrre}{{{\ifmmode {\cal R}_3^{E0}}\else ${\cal R}_3^{E0}$\fi}}
\newcommand{\rrrp}{{{\ifmmode {\cal R}_3^P}\else ${\cal R}_3^P$\fi}}
\newcommand{\jdde}{{{\ifmmode {\cal D}_2^{E0}}\else ${\cal D}_2^{E0} $\fi}}
\newcommand{\jddp}{{{\ifmmode {\cal D}_2^P}\else ${\cal D}_2^P$\fi}}
\newcommand{\ycut}{{{\ifmmode y_{cut} }\else $y_{cut}$\fi}}
\newcommand{\ymin}{{{\ifmmode y_{min} }\else $y_{min}$\fi}}
\newcommand{\sph}{{{\ifmmode {\cal S} }\else ${\cal S} $\fi}}
\newcommand{\apl}{{{\ifmmode {\cal A} }\else ${\cal A} $\fi}}
\newcommand{\thr}{{{\ifmmode {\cal T} }\else ${\cal T} $\fi}}
\newcommand{\obl}{{{\ifmmode {\cal O} }\else ${\cal O} $\fi}}
\newcommand{\cpa}{{{\ifmmode {\cal C} }\else ${\cal C} $\fi}}
\newcommand{\eec}{{{\ifmmode {\cal E}{\cal E}{\cal C} }\else
${\cal E}{\cal E}{\cal C}$\fi}}
\newcommand{\aeec}{{{\ifmmode {\cal A}{\cal E}{\cal E}{\cal C} }
\else ${\cal A}{\cal E}{\cal E}{\cal C}$\fi}}
\newcommand{\hjm}{{\ifmmode {\bf M^2_{high}}
                   \else   ${\bf M^2_{high}}$\fi}}
\newcommand{\ljm}{{\ifmmode {\bf M^2_{low}}
                   \else   ${\bf M^2_{low}}$\fi}}
\newcommand{\djm}{{\ifmmode {\bf M^2_{diff}}
                   \else   ${\bf M^2_{diff}}$\fi}}
\newcommand{\hjmt}{{\ifmmode {\bf M({\cal T})^2_{high}}
                    \else   ${\bf M({\cal T})^2_{high}}$\fi}}
\newcommand{\ljmt}{{\ifmmode {\bf M({\cal T})^2_{low}}
                    \else   ${\bf M({\cal T})^2_{low}}$\fi}}
\newcommand{\djmt}{{\ifmmode {\bf M({\cal T})^2_{diff}}
                    \else   ${\bf M({\cal T})^2_{diff}}$\fi}}
\newcommand{\djr}{{{\ifmmode {\bf {\cal D}_2}\else ${\bf {cal D}_2}$\fi}}}
\newcommand{\ma}{{{\ifmmode {\bf {\cal M}_{Major}}
\else ${\bf {\cal M}_{Major}}$\fi}}}
\newcommand{\mi}{{{\ifmmode {\bf {\cal M}_{Minor}}
\else ${\bf {\cal M}_{Minor}}$\fi}}}

\newcommand{\ha}{{{\ifmmode {\frac{1}{2}}\else ${\frac{1}{2}}$\fi}}}

\renewcommand{\arraystretch}{1.7}

\begin{document}
\begin{titlepage}

\begin{flushright}
IEKP-KA/98-16 \\[3mm]
{\tt hep-ph/9808448}
\end{flushright}
\vspace{4.cm}

\begin{center}
  {\large\bf Search for SUSY and Higgs particles\footnote{Invited
      talk at the XVIII Physics in Collision Conference, Frascati, 16-20 June 1998.}}\\[1cm]

  {\bf W.~de Boer} \\[5mm]
  {\it Institut f\"ur Experimentelle Kernphysik, \\University of Karlsruhe \\
       Postfach 6980, D-76128 Karlsruhe, Germany} \\[1cm]
\end{center}

\vspace{2cm}

\abstract{

An overview of hints for new physics outside the Standard Model
and the status of  sparticle and Higgs searches is given.
The present limits on Higgs bosons of about 90 GeV as well
as the $\bsg$ rate and cosmological constraints severely
restrict the available parameter space of the MSSM.
}

\end{titlepage}

%
\baselineskip=17pt

%
%
%
\section{Introduction}

Interest in supersymmetry, the symmetry between
fermions and bosons, originated from the fact, that
it is a non-trivial extension of the Poincar\'e group,
 which now includes 
internal quantum numbers of particles, thus paving
the way for a unification of   strong, electromagnetic
and weak interactions with gravity.
In addition, supersymmetry removed the ultraviolet divergencies
that plagued the Standard Model (SM). Details about
these developments in the 1970s can be found in many reviews\cite{rev}.

Interest in supersymmetry
became a big boost in the 1990s after precise measurements
 of the gauge couplings at LEP, which showed that
 gauge coupling unification, the prerequisite
for  Grand Unified Theories (GUT), is only possible
in the supersymmetric
extension of the SM, not in the SM itself\cite{unif}.
However, the price to be paid for this symmetry
is a doubling of the particle spectrum\cite{rev}:
for each fermion (boson) of the SM one needs to introduce
an additional boson (fermion) with the same quantum numbers.
In addition, two Higgs doublets instead of one doublet are
required. This minimal supersymmetric extension is called
the Minimal Sypersymmetric Standard Model (MSSM).

Supersymmetry cannot be  an exact symmetry in nature
as  the  superpartners must be heavier than
the SM ones: none of the
predicted spin 0 partners of the quarks and leptons,
the so-called squarks and sleptons, nor the spin 1/2 partners
of the gauge bosons, the photino, zino, wino and gluino,
have been observed so far. It should be noted that some of
the higgsinos and gauginos
have the same quantum numbers, such as spin  and
electric charge, thus
allowing  mixing between the mass and interaction eigenstates:
 the mixed states of the wino and charged higgsino are usually
 called chargino whilst the mixed states of
 the photino, zino and two neutral higgsinos
 are the so-called neutralinos.
 The  detailed properties of the mass mixing matrices
 and mass relations
can be found in standard reviews\cite{rev}.

In addition to gauge unification at a scale $\mgut\approx 10^{16}$
GeV, the Yukawa couplings of the $b$-quark and $\tau$-lepton
turned out to unify at the same scale in the MSSM,
as expected in practically
all GUT's, since the  quarks with charge (--1/3) and leptons
belong to  the same representation of any group
containing the well-known $SU(3)\otimes SU(2)\otimes U(1)$ from
the SM as subgroups\cite{rev}.

After the discovery of the heavy top quark the Higgs mechanism
has a natural explanation in the MSSM because of  the large radiative
corrections
from the Yukawa couplings to the Higgs potential, which
can easily  introduce  a non-trivial minimum due
to the difference between the running of the masses
of the two Higgs doublets\cite{rev}. Therefore,  the electroweak
symmetry breaking  need not  be introduced
{\it ad hoc}, as in the SM, but its origin is the heavy top
quark, thus linking intimately the $Z^0$ mass and the top mass.
This link only functions for $140 < m_t < 200$ GeV and therefore
the experimental top mass is exactly in the range
required by the MSSM.

Furthermore, the coupling in the
Higgs potential is not arbitrary, as in the SM, but
constrained by the gauge couplings. This allows the prediction
that the Higgs mass will be below 130 GeV (preferentially
even below 100 GeV). Such a low Higgs mass is indeed
preferred by the electroweak precision data\cite{ewwg},
as will be discussed in Section \ref{hints}.
The direct observation of a Higgs mass in the predicted range
would certainly provide another big  boost for the MSSM, although
its ultimate verification can only come from the
direct proof  of the existence of sparticles.

During recent years' several deviations from the SM
have been suggested as hints of supersymmetry.
Although none of the few sigma deviations themselves
were convincing, it has been argued that most
of them pointed to the same region of the  MSSM parameter
space\cite{kane}.
However, most of these `hints' have {faded away during the
last year, as will be discussed in the next Section.

As long as  the origin of the breaking
of supersymmetry is not known, one has to search for
direct signs of SUSY
in very diversified ways. Among the models obtaining  most attention
for  direct searches with the present colliders are:
{
\begin{itemize}
    \item {\underline{The \rpv ~scenarios.}}\\
    In the SM the decay of a quark into a lighter quark plus lepton
    is effectively suppressed by angular momentum conservation
    since all have spin 1/2. However, in SUSY
    this is not the case, so quarks can e.g. decay into a
    lepton and a squark, leading to 
     Baryon- and Lepton-number violation, and consequently to
     proton decay, if no precautions are
    taken. To avoid such B- and L-violating interactions
    one usually assumes that
    the multiplicative quantum number $R_p=(-1)^{3B+L+2S}$ is
    conserved. This R-parity is +1 for SM particles and --1 for
    SUSY particles. As a result of $R_p$ conservation, sparticles
    have to be produced in pairs and  the decay products
    of any SUSY particle ($R_p=-1$)  must contain another SUSY
    particle, thus for kinematical reasons the lightest
    SUSY particle cannot decay
    anymore  and must therefore
    be stable.
     These properties define it as the perfect candidate
    for the dark matter in our universe, provided it is neutral too,
    which is the case in many scenarios.
    The non-interacting stable  LSP  leads to the
    famous missing energy  signatures for supersymmetry.
    If R-parity is broken, there is no missing energy and momentum,
    but the SUSY signature would consist of events
    with many jets and/or multiple leptons\cite{rpv}.
 \item {\underline{The gauge mediated scenario.}}\\
   In the gauge mediated scenario
   the breaking of supersymmetry is caused by gauge interactions.
    Since the breaking is proportional to the
    gauge couplings, one expects the  SUSY masses
    to be proportional to the gauge couplings times a breaking scale,
    i.e.
    $M_{\tilde{B}}\approx(\alpha_1/4\pi)\Lambda^2_{SUSY}$.
        For SUSY masses in the 100 GeV to 1 TeV range,
   the  breaking scale $\Lambda^2_{SUSY}$ has to be of the order of
   $10^2-10^3~ \rm TeV$.
   In such scenarios the gravitino
 $M_{\tilde{G}}\approx ({\Lambda^2_{SUSY}}/{M_{Pl}})$
  is of the order of a few eV, which implies that  it is the LSP,
  but  it is too light to be a  candidate for cold dark matter.

  The  sparticles can decay into  a gravitino plus a photon.
    The famous CDF event\cite{cdf} with two isolated charged particles
   and two photons was the perfect candidate for such
   a scenario (see  Section \ref{hints}) and consequently
    many searches, both at LEP and the Tevatron, concentrated
   on  final states with isolated photons,
   albeit without succes so far\cite{cdf,susywg,d0}.
      \item \underline{The supergravity inspired scenario.}\\
      If the symmetry breaking is due to flavour blind gravitational
      interactions, one usually assumes a common mass
      ($m_0$) for the scalars
      and one ($m_{1/2}$) for the spin 1/2 gauginos at the GUT scale.
      Owing to radiative corrections the masses become
      different at the electroweak scale, but in a well
      defined manner given by 
      Renormalization Group Equations (RGE).
 \end{itemize}    
}

Space does not permit  all scenarios to be covered exhaustively.
I will concentrate on the supergravity inspired
scenario, which is the most interesting one for cosmology,
since it provides a very natural candidate for dark matter
for a large region of parameter space, as will be discussed
in  Section \ref{constraints}.

 The status of the
present sparticle and Higgs searches will be summarized
in Sections \ref{susy} and \ref{higgs}, respectively.
The present  limits  severely constrain the parameter space of the
MSSM, as will be discussed in Section \ref{constraints}.

%
\section{Hints for SUSY?}
\label{hints}
In this section, we discuss the status of several deviations
from the SM which have been suggested as possible hints
of supersymmetry.
\subsection{The CDF $ee\gamma\gamma\etmiss$ event.}

Several years ago, the CDF collaboration found  an interesting
event with two isolated electrons and two isolated energetic photons
plus missing energy\cite{cdf}. Since  the probability
from SM  backgrounds
(mainly radiative W-pair production) was low $\approx 10^{-6}$,
the  origin
might be selectron pair
production with each
selectron decaying into an electron plus an unstable photino-like
neutralino.
The latter one can
decay either into a photon plus Higgsino-like LSP\cite{kane1} or
a photon plus a light gravitino\cite{dimo}.
A light gravitino  is expected in gauge-mediated supersymmetry-breaking
scenarios, the Higgsino-like LSP in supergravity
inspired scenarios with a small Higgs mixing parameter $\mu$.

Recently, the CDF Collaboration  reanalysed and
published the event\cite{cdf}.
They concluded that after the final detector alignment
one of the tracks, previously defined as an electron,
did not lign up with the electromagnetic cluster anymore.
In addition, the invariant mass between track and cluster
was above the $\tau$ mass, so  the initial lepton interpretation
is doubtful.

Furthermore, if one of the SUSY interpretations was
correct, one would expect 
more events from other SUSY processes in the inclusive
$\gamma\gamma$ sample, i.e. without the requirement of
lepton tagging. None were found by CDF in the full data sample
of about 90 $\rm pb^{-1}$. This search was used
to set upper limits on chargino production\cite{cdf}.
A similar search by D0 was used to set limits on squark
production\cite{d0}.

\subsection{HERA anomalies}

From the 1994--1996 data, both H1 and ZEUS reported
anomalies in the high $Q^2$ region\cite{meyer}.
The clustering of the H1 events at an invariant mass around
200 GeV led to speculations about the s-channel resonance
production of a leptoquark, which would occur naturally
in squark decays, if R-parity is violated.
However, for the 1997 data\cite{meyer}, in which the luminosity
was more than doubled,
the number of events was lower than, but compatible with the
expectation in the region of interest and one
should  consider the anomalies in the 1994--1996 data
as statistical fluctuations.

\subsection{ALEPH 4-jet anomaly}
\begin{figure}[t]
   \epsfig{file=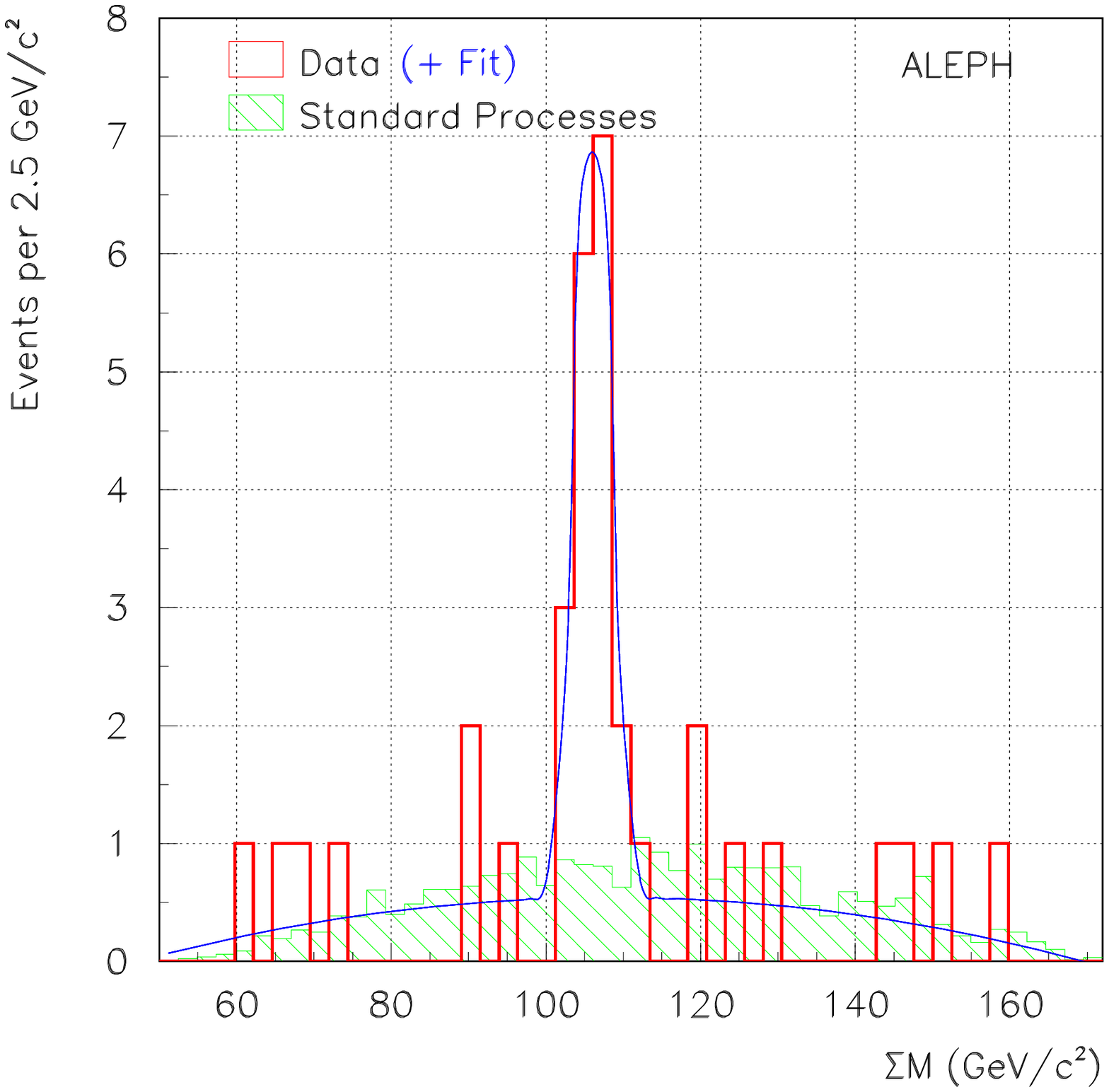,width=0.48\textwidth}
 \epsfig{file=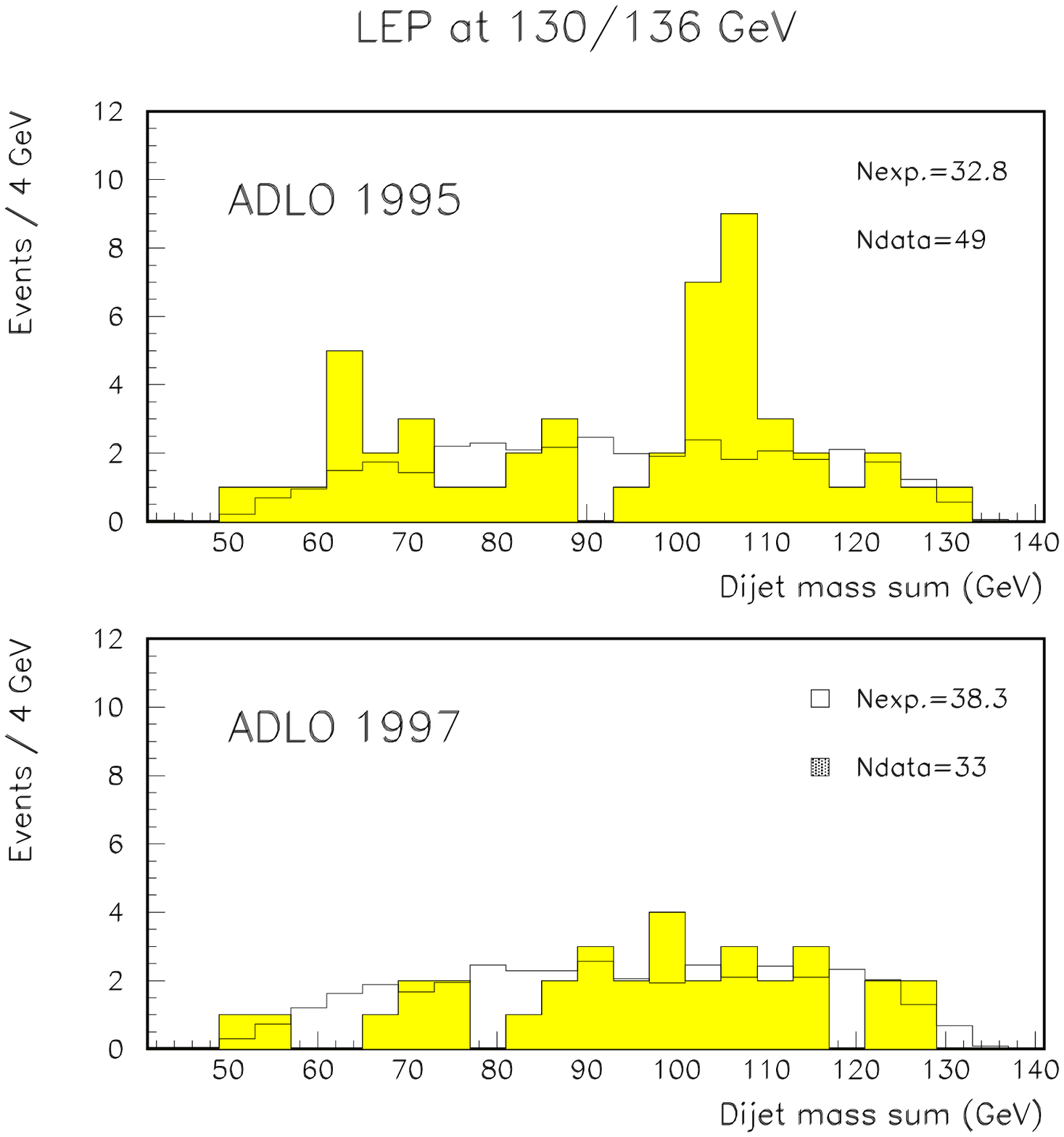,width=0.50\textwidth}
\caption{\label{aleph1} \it
Left:  The ALEPH 4-jet anomaly from all data presented
  at the LEPC Meeting in November 1996.
    Assuming the production of two new objects
  X and Y with similar masses and each decaying into 2 jets,
  one obtains
  the best resolution in the plotted sum of $M_X$ and $M_Y$. 
  Jet-pairing is chosen such that the mass difference between
  the pairs is minimized. Right: the 4-jet data of all four LEP
  experiments for the 135 GeV runs in 1995 and 1997. The excess
  in 195 GeV was dominated by the ALEPH events, but could not
  be reproduced in 1997.}
\end{figure}
ALEPH discovered  a
      splendid signal
    during  their  Higgs search in four jets\cite{aleph}, when {\it no}
     b-tagging was required,
         as shown in Fig. \ref{aleph1}.
Possible MSSM interpretations included  the pair production
of left and right handed selectrons\cite{wagner} or charginos\cite{pok}.
However, the results were not confirmed by the other
experiments. A working group of the four experiments
concluded that all experiments had a similar efficiency
for 4 jets, so it was unclear why it only showed up in one
experiment.
Finally, the LEPC Committee decided to have a new run
at 135 GeV, where the bulk of the ALEPH signal had accumulated.
However, after two weeks of running the total accumulated
luminosity of 24 $\rm pb^{-1}$ from the four experiments did
not show any signal (see Fig. \ref{aleph1}), and therefore this
$\approx 7-10~\sigma$ ALEPH anomaly
should  only be remembered as an anomaly,
 not new physics.
\begin{figure}[t]
\epsfig{file=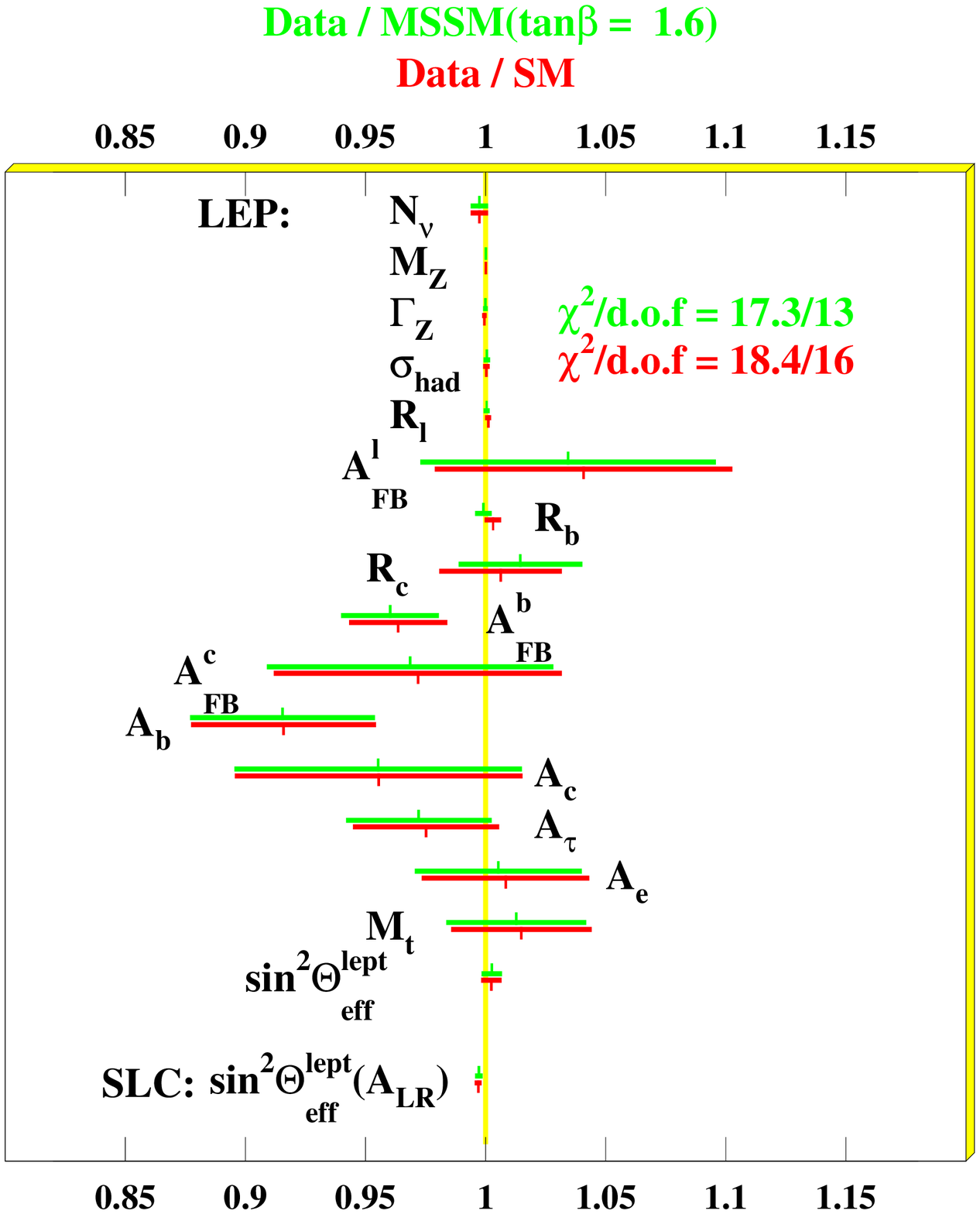,width=.47\textwidth}
\epsfig{file=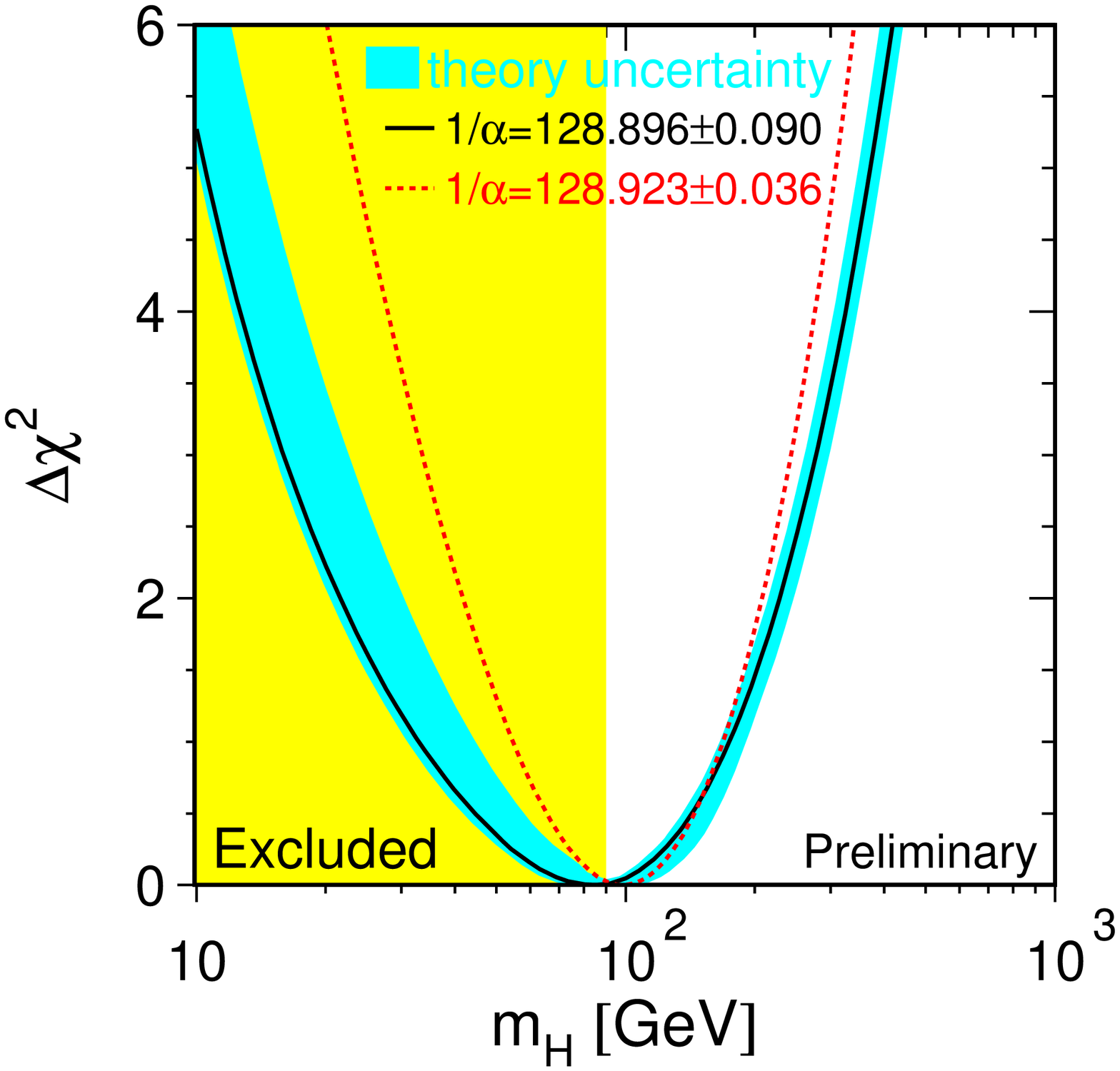,width=.51\textwidth}
\caption[]{\it
 Left:    Comparison of ratio of the electroweak measurements and
     theory in the SM and MSSM. Right:
      the $\chi^2$ distribution as function of the Higgs mass
      from the SM fit to the electroweak precision
      observables and the top mass.
The shaded area is excluded by the direct searches.
    \label{balken} }
\end{figure}
%
\begin{figure}[t]
  \vspace*{-0.8cm}
\epsfig{file=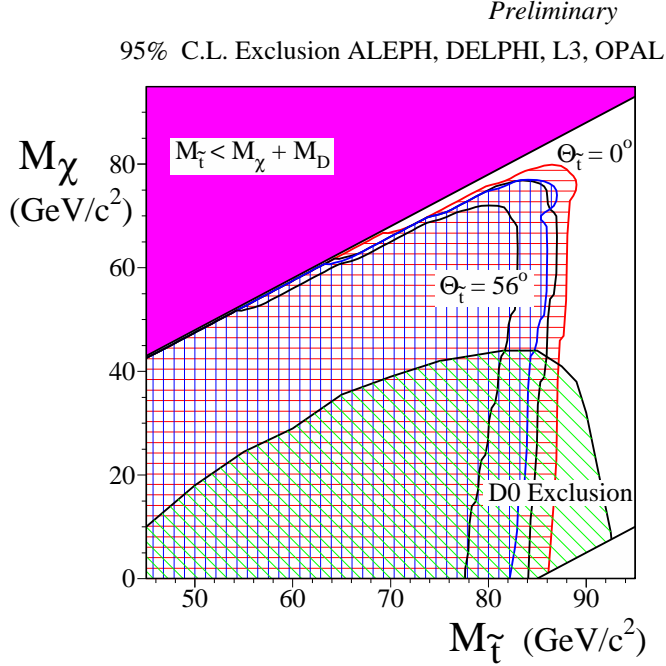,width=.59\textwidth}
\epsfig{file=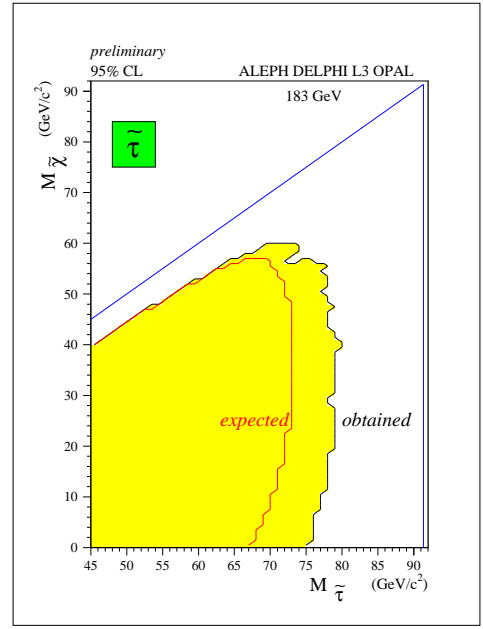,width=.39\textwidth}\vspace{-2mm}
\caption{\label{leplim}\it
 Excluded  right-handed stop and stau masses as a function of the
 LSP mass. For the stop the excluded region by the D0 Collaboration
 is shown
  as well as the pessimistic case of a mixing angle of $56^o$
 between left- and right-handed stops, which minimizes the
 coupling to the $Z^0$ boson.  For the stau the expected limit,
 based on the expected number of events instead of the observed
 number of events, is also shown.  }
\end{figure}
\begin{figure}[t]\vspace*{-0.5cm}
\epsfig{file=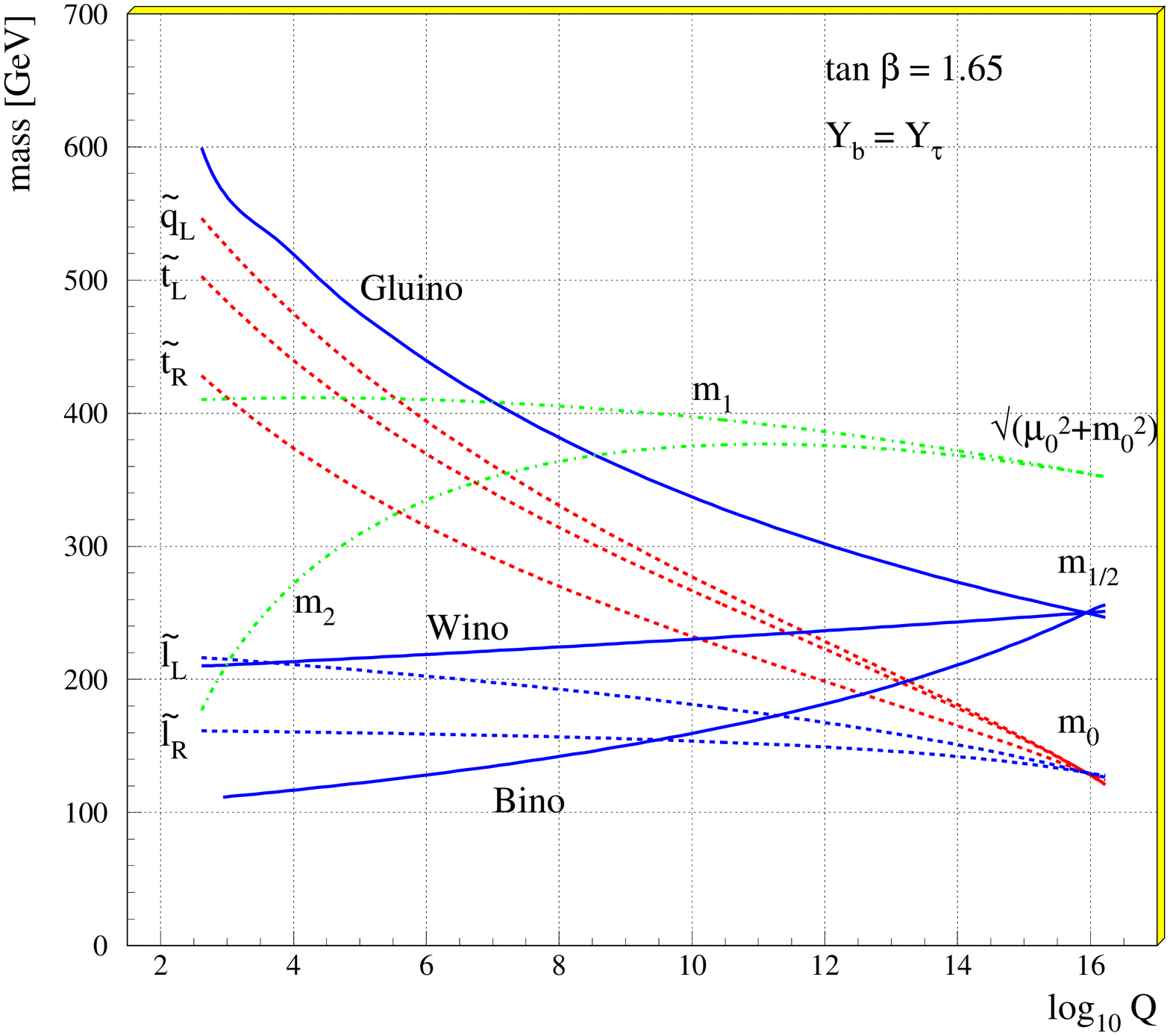,width=.48\textwidth}
\epsfig{file=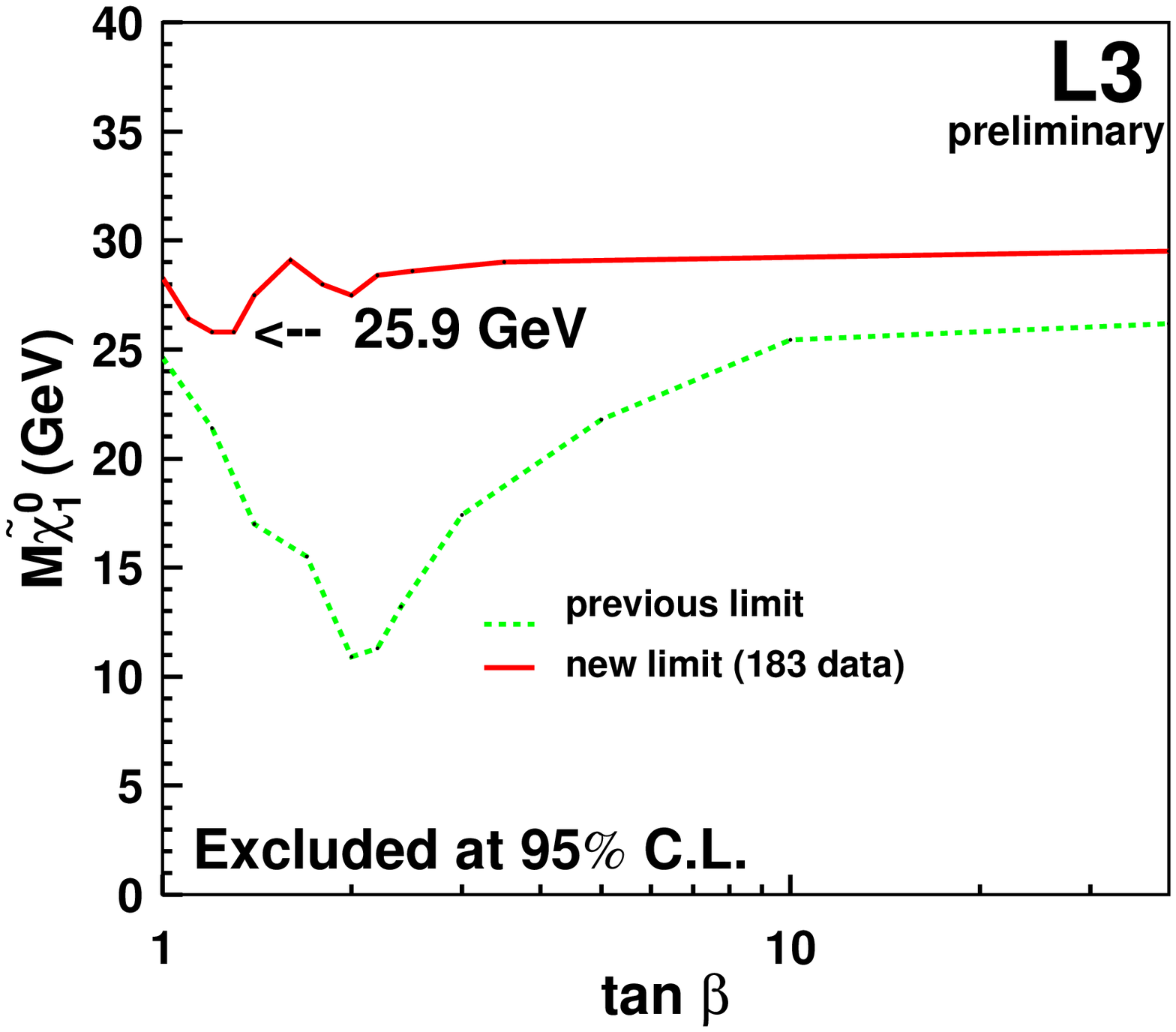,width=.48\textwidth}
\caption{\it
 Left: Evolution of SUSY masses. Note that in the CMSSM: a) the LSP is the bino-like neutralino;
 b) the  right-handed sfermions are lighter than the left-handed ones;
 c) the LSP is roughly half the chargino mass at low $\tb$;
 d) the pseudoscalar Higgs mass, $\mA=m_1^2+m_2^2$+rad. corrections,
  is usually large, both because $m_0$ and $\mu$ are large ($m_0$ because of the  relic density constraint,
  $\mu$ because of electroweak symmetry breaking, see section \ref{constraints}).
 Right: Limit on the LSP as a function of $\tb$ for any value of $\mze$.
    \label{lsplim} }
\end{figure}
\subsection{Electroweak precision measurements}
The MSSM can describe  Electroweak Precision Measurements (EPM)
at least as well as the SM, as demonstrated in Fig. \ref{balken}.
These results are an update from Ref.\cite{hollik1} with the
summer 1998 data, as compiled
by the Electroweak Working Group\cite{ewwg}. 
The famous deviation in $R_b=\Gamma(Z^0\rightarrow b\overline{b})/
\Gamma(Z^0\rightarrow hadrons)$ has gradually come down
from a 3.5 $\sigma$ deviation a few years ago, to a 1.4 $\sigma$
effect at present. The better agreement in the MSSM for this
variable mainly stems from  charginos-stop corrections to the b-quark
production vertex, although this requires light
 stop and chargino masses. The improvement with the 
present exclusion limits for stops and charginos above 90 GeV
(see Section \ref{susy})
is small.

Another interesting result from electroweak fits is
the fact that they point to a low Higgs mass\cite{ewwg},
as  is apparent from the $\chi^2$ distribution in Fig. \ref{balken}.
However, the discrepancy between LEP and SLC on $\sws$ is
still a dominant source of uncertainty in the Higgs mass:
$\sws=0.23101\pm 0.00031$ from SLC\cite{schwartz} corresponds
to a Higgs mass of $m_H=17^{+27}_{-8}$ GeV, while the LEP value
of $\sws=0.23183\pm0.00021$ corresponds to $m_H=176^{+120}_{-76}$ GeV.
The combined value is  $m_H=100^{+72}_{-45}$ GeV.
If the combined fit is performed with the requirement that
the Higgs mass has to be above 90 GeV, the SLC data get less weight,
thus leading to a  higher upper limit on the Higgs mass\cite{chan,jer}.

\section{Search for SUSY particles}\label{susy}
\begin{figure}[t]\vspace*{-5mm}
  \begin{center}
   \epsfig{figure=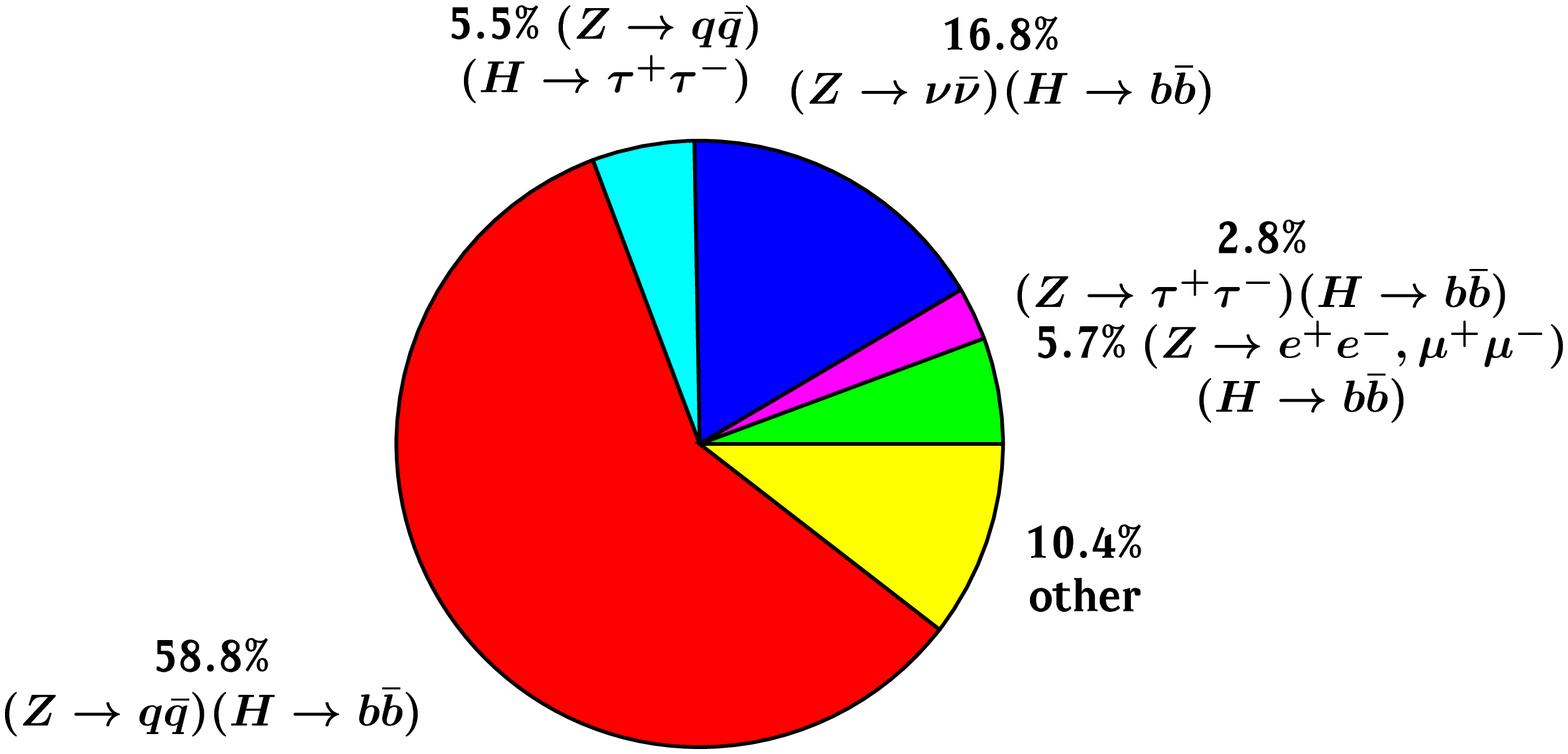,width=0.26\textwidth,angle=90}
      \epsfig{figure=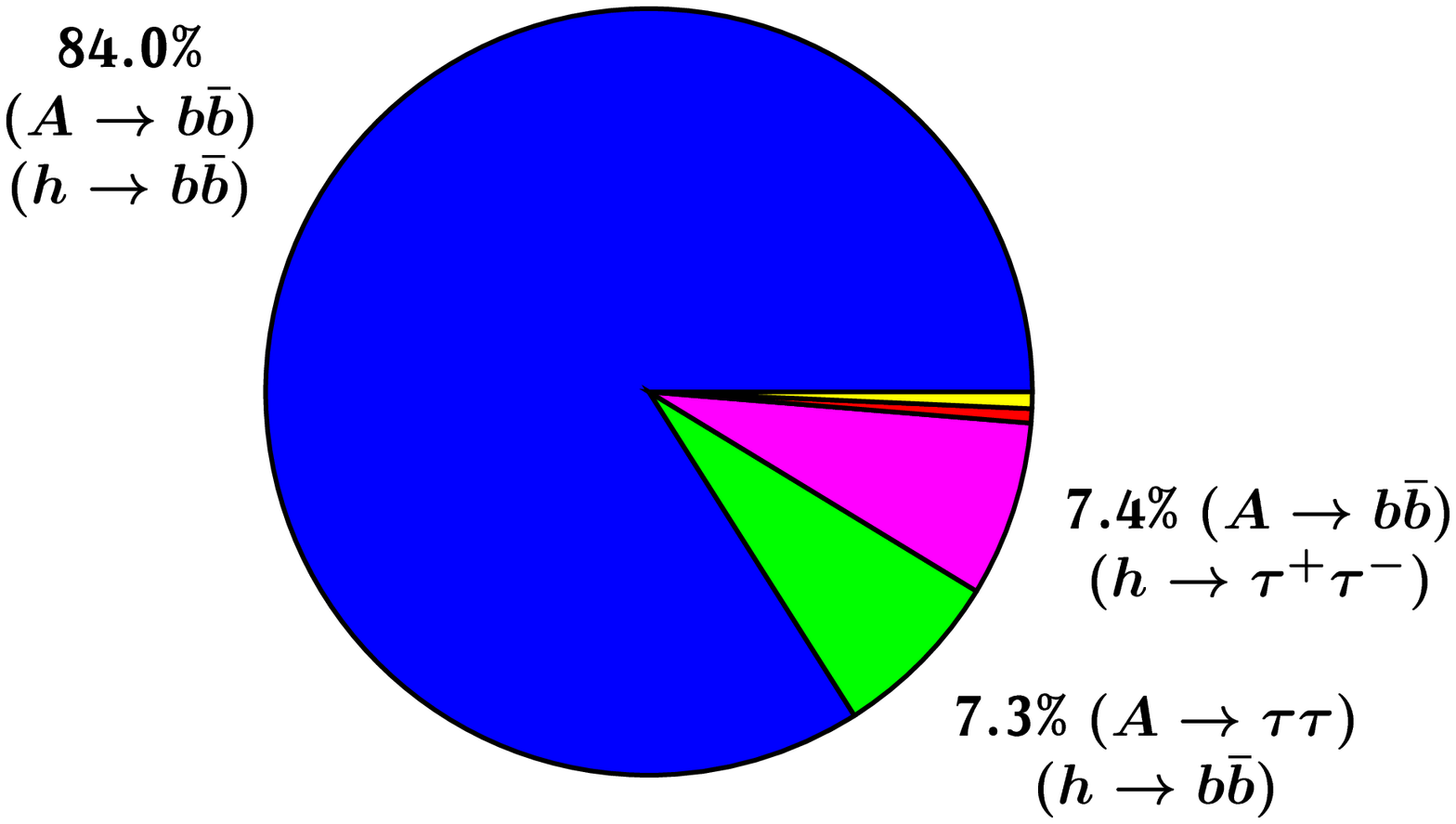,width=0.31\textwidth,angle=90}
\caption[]{ \it Left:
Final states in Higgs production ($\mh=85$ GeV, $\sqrts=183 \rm{GeV}$)
from hZ Bremsstrahlung. Right: as on the left-hand side
for hA production.
\label{pie}
}
\end{center}
\end{figure}

As mentioned in the introduction, we  restrict ourselves
here to the supergravity inspired scenarios with common
mass scales at the GUT scale and a stable LSP.
In these scenarios, the signature is the missing energy and momentum.
However, backgrounds from $\gamma\gamma$, $We\nu_e$, $Zee$,
$WW$ and $ZZ$ production have to be suppressed by suitable
cuts on variables like visible energy, visible mass, thrust, etc.
Typically, the number of candidates after the cuts
can be reduced to a few events. Up to now, the
number of candidates is consistent with the background,
so one can set only upper limits on the SUSY cross sections
for a given mass, or alternatively, these cross section limits
can be transformed into mass limits.
These limits  depend on the LSP mass, since it defines
the amount of missing energy, which is an important
criteria to separate the signal from the background
(assuming that  enough energy is seen in the detector to
trigger the event).
If the mass difference $\Delta M$ between the LSP and
the SUSY sparticle
mass is at least 15 GeV, the background is usually no problem.
If $\Delta M$ becomes smaller, the visible energy rapidly
decreases and the background, especially 
the $\gamma\gamma$ background,
rapidly increases.
If $\Delta M$ becomes only a few GeV, the lifetime of the
sparticles can become so long that they decay inside or even
outside the detector, thus forming kinks in the charged tracks
or resembling stable charged particles. The results of such scenarios
have been summarized by the SUSY Working Group\cite{susywg}.
Therefore, it is useful to give  upper limits on the
cross sections in a plane of the sparticle mass versus the LSP mass.
 Such plots are shown in Fig. \ref{leplim} for  stops and staus,
 which are expected
 to be the lightest sfermions due to the negative
 corrections from the Yukawa couplings.
  These preliminary plots  were
 prepared by the SUSY Working
 Group\cite{susywg}.

The upper limits on the cross section can be transferred
into lower limits on the sparticle masses, which are
 indicated in Table \ref{t2} together with typical
expected masses in the Constrained MSSM (CMSSM), which will be
discussed in 
Section \ref{constraints}.
These limits are only valid for a sufficiently large $\Delta M$,
since if the LSP mass is close to the sfermion mass, the
visible energy in the detector is too small. This  can be observed
from the small unexcluded regions close to the diagonal in
Fig. \ref{leplim}. Note that these regions are much larger
for hadron colliders, as shown by the D0 exclusion plot in
Fig. \ref{leplim}.

 LSP production yields only invisible final states,
except for initial state radiation. Nevertheless,
LSP mass limits can be obtained, if one assumes unified gaugino masses at the GUT scale, which yield  mass relations between neutralinos and charginos at low energies (see left-hand side of Fig. \ref{lsplim}).
Using these assumptions, L3 finds a
lower limit
of 25.9 GeV for the LSP mass for low $\tb$ and any value of
$m_0$ allowed by the slepton limits\cite{l3}.
For high values of $\tb$ and $\mze$
the LSP limit is about half the chargino limit, which follows
directly from the evolution of the masses by the RGE.
For large $\mze$ the sneutrinos are heavy and the chargino
cross sections is not decreased
by the negative interference from the t-channel sneutrino exchange.
Other experiments
have performed similar analysis with similar results\cite{lsplim}.
\section{Search for Higgs bosons}\label{higgs}
\begin{figure}[t]
  \begin{center}
    \epsfig{figure=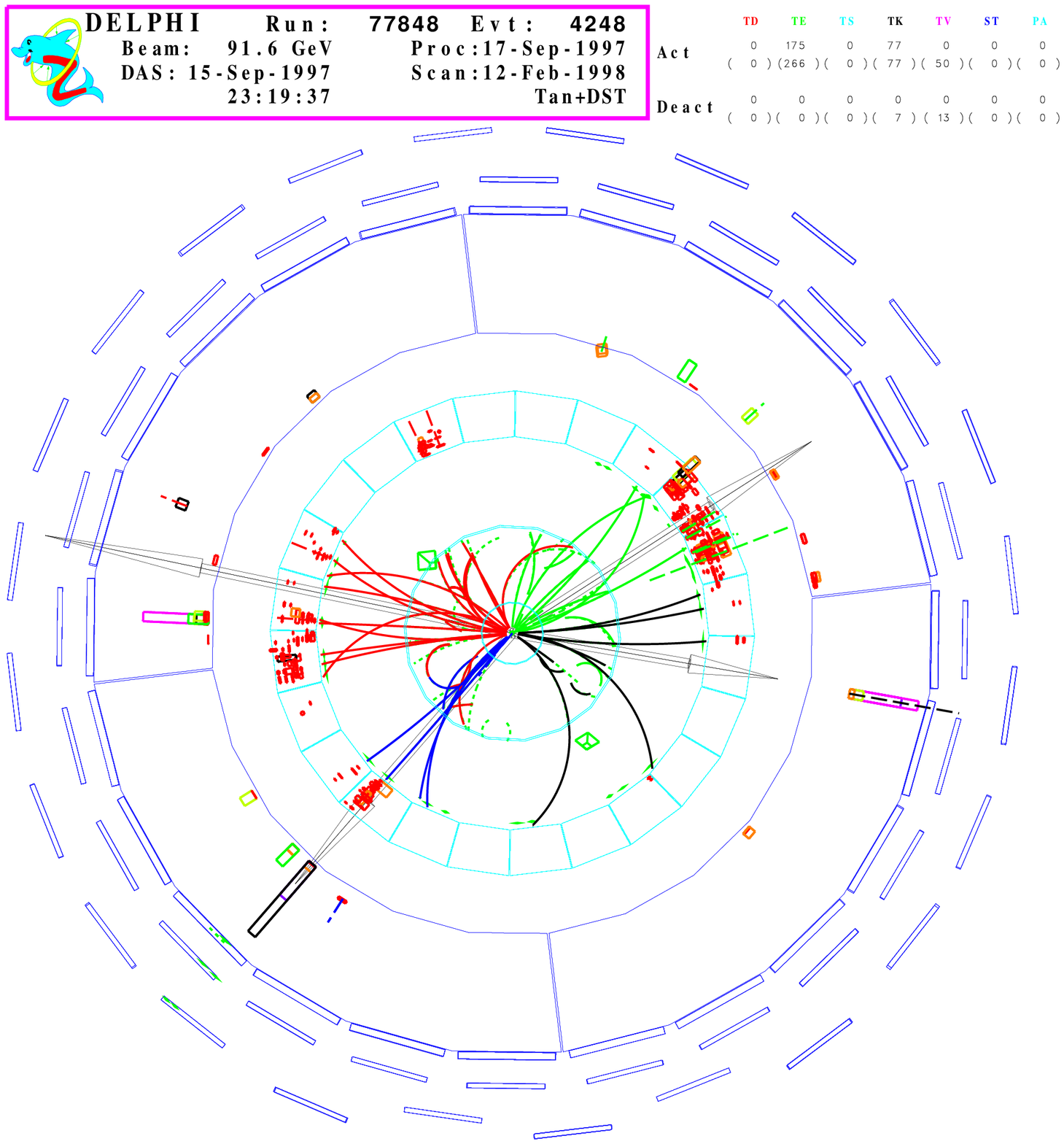,width=0.3\textwidth}
    \epsfig{figure=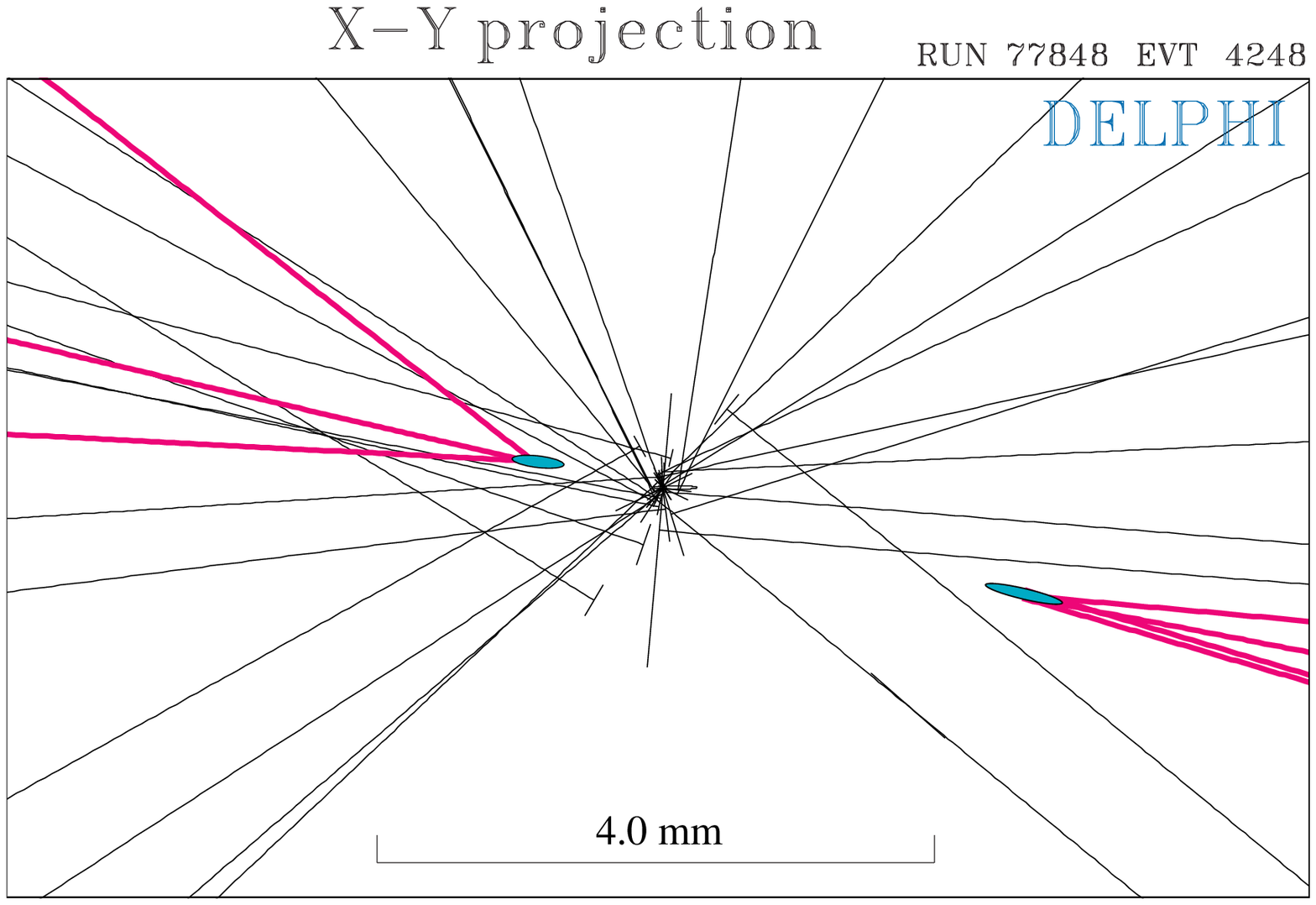,width=0.44\textwidth}
\caption[]{ \it
DELPHI Higgs candidate in the $b\overline{b}qq$ channel.
Note the large secondary vertex, characteristic of the decay of
a B-meson, in the blown up version of the vertex region
on the right-hand side.
The invariant masses of the quark pairs
are 89 and 91 {\rm GeV}, respectively, so the event is consistent with
a ZZ candidate.\label{aleph}}
\end{center}
\end{figure}

\begin{figure}[t]\vspace*{-0.8cm}
\begin{center}
\epsfig{figure=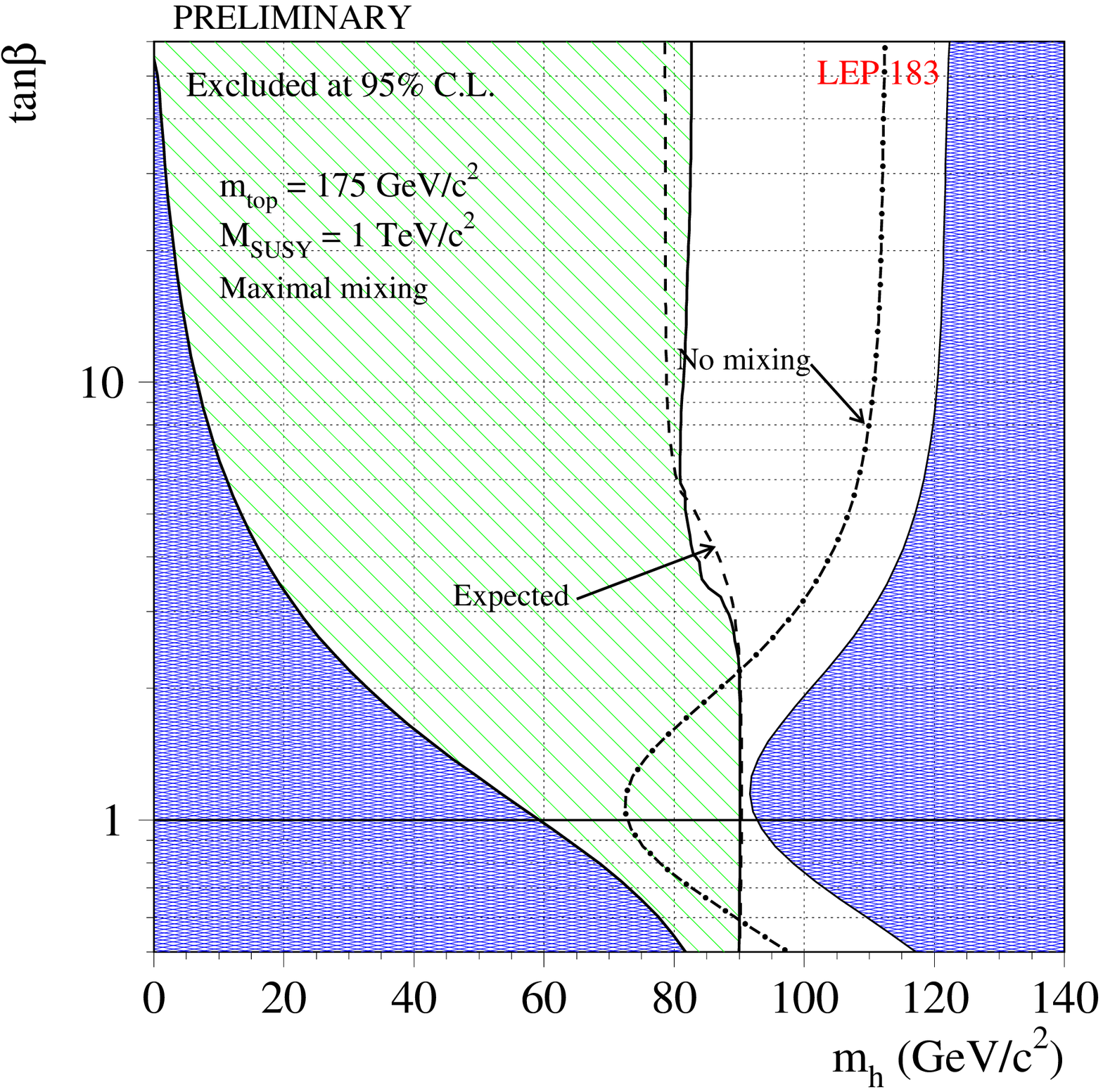,width=0.47\textwidth}
\epsfig{figure=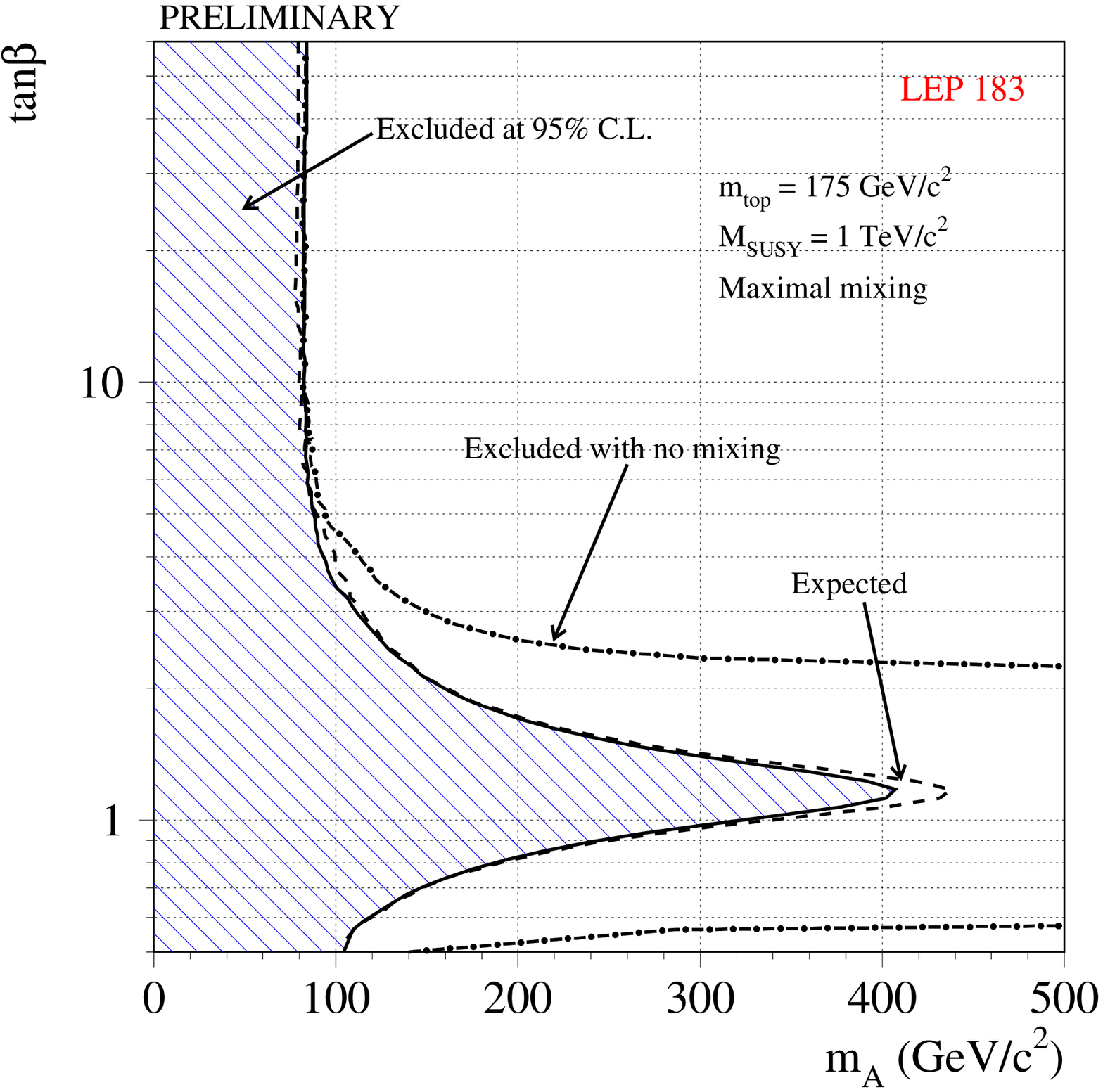,width=0.47\textwidth}
\caption[]{ \it
  Combined 95\% CL exclusions in the (\mh, \tanb)  (left)
  (\mA, \tanb) (right) plane.
  The corresponding expected limit
   is also shown (dashed curve). 
  The dark regions are not allowed by theory. Here the
  most conservative case of maximal 
   mixing between the left- and right-handed
   stop squarks  was assumed. The other extreme of no mixing
   is indicated by the dashed-dotted curves.
\label{limits}}
\end{center}
\end{figure}
  \begin{figure}[t]
    \begin{center}
\vspace*{-1.8cm}
    \epsfig{file=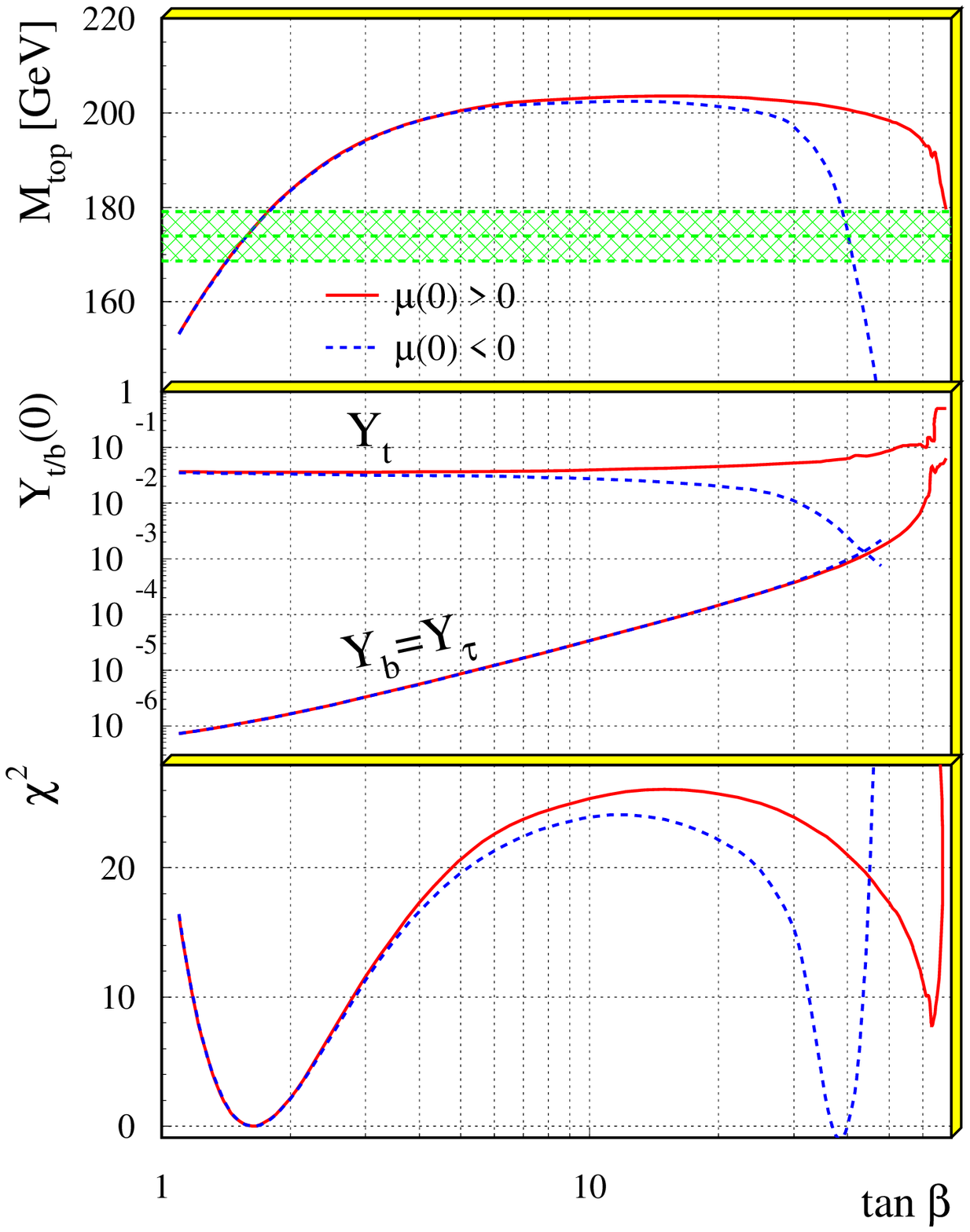,width=0.41\textwidth}
  \epsfig{file=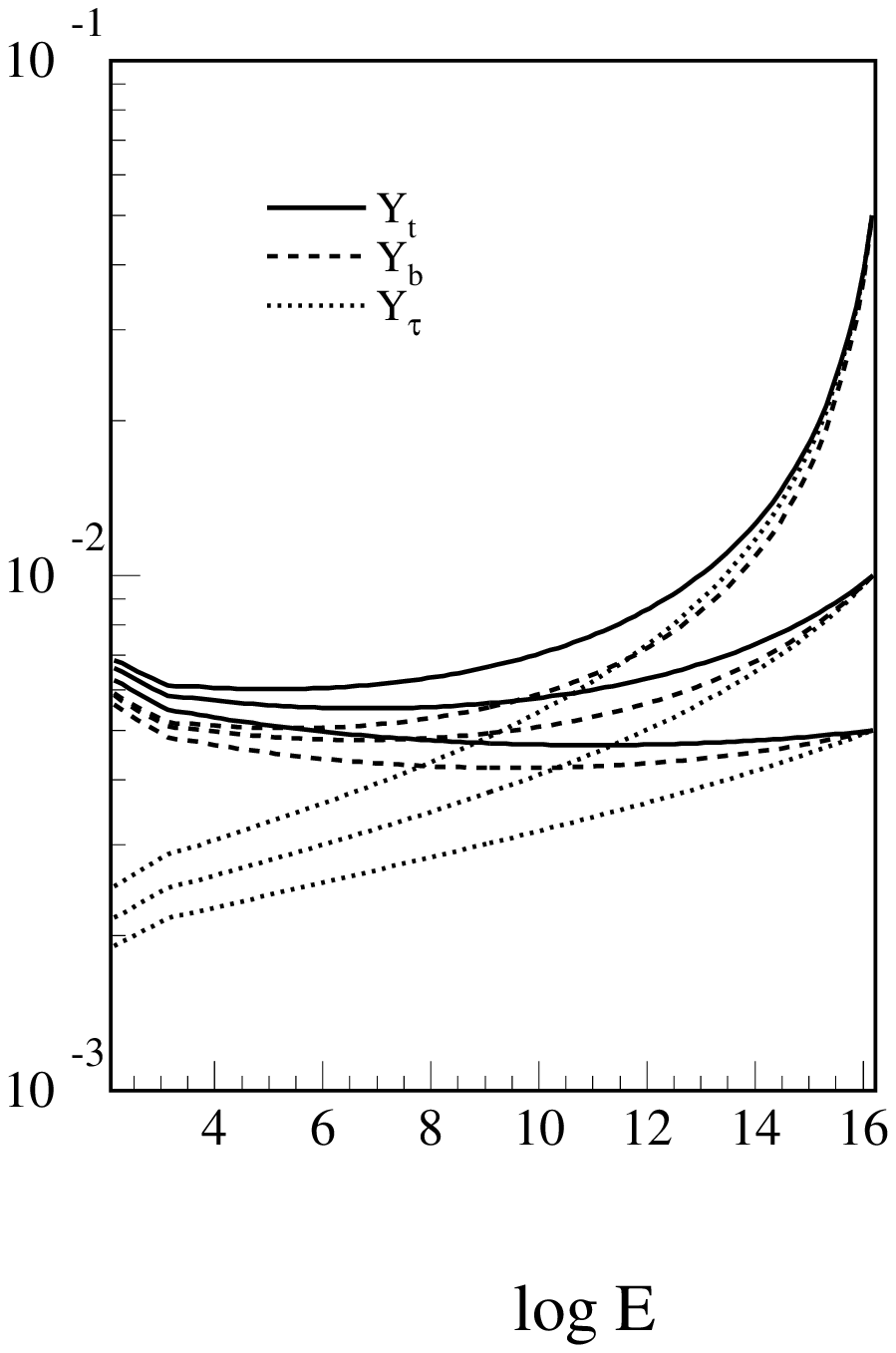,width=0.43\textwidth}
\end{center}
\vspace{-.5cm}
  \caption[]{\label{f2}\it Left: The top quark mass as a
    function of $\tb$ (top) 
      for values of $\mze,\mha~\approx 1 \rm{TeV}$
      after requiring $b-\tau$ unification.
      The curve is hardly different 
       for lower SUSY masses.
      The middle part shows the corresponding values of the Yukawa
      couplings at the GUT scale and the lower part the
      $\chi^2$ values.
      If the top constraint ($\mt=174\pm5$, horizontal band) 
      is not applied, all values of $\tb$ between 1.2 and 
      70 are allowed,
       but if the top mass 
      is constrained to the experimental value, only the regions
      $\tb=1.65\pm0.3$, $ \tb \sim 35$, and $ \tb \sim 64$  are 
      allowed. Right: The running of the Yukawa couplings when
$Y_t=Y_b=Y_\tau$ at the GUT scale ($SO(10)$ type solution).
One can clearly 
 see the approach to the three different fixed points,
 i.e. the
value at low energy is largely independent of the GUT scale value.
Consequently the
GUT scale values can be chosen to be equal (triple unification).
The fixed point values at low energy  yield
 correct masses for   bottom and tau for $\tb\approx 64$; the fixed
point of the top mass yields 
$m_t^2\equiv (4\pi)^2 Y_t v^2 \sin^2\beta$=184 {\rm GeV}, 
which is about $2\sigma$ above
the experimental value.  At low $\tb$ only $Y_t$ is large (see
left-hand side),
in which case also $Y_t$ shows an infrared fixed point behaviour.
      }
 \end{figure}
%
%
%
  \begin{figure}[t]
    \vspace*{-.7cm}
\begin{center}
    \epsfig{file=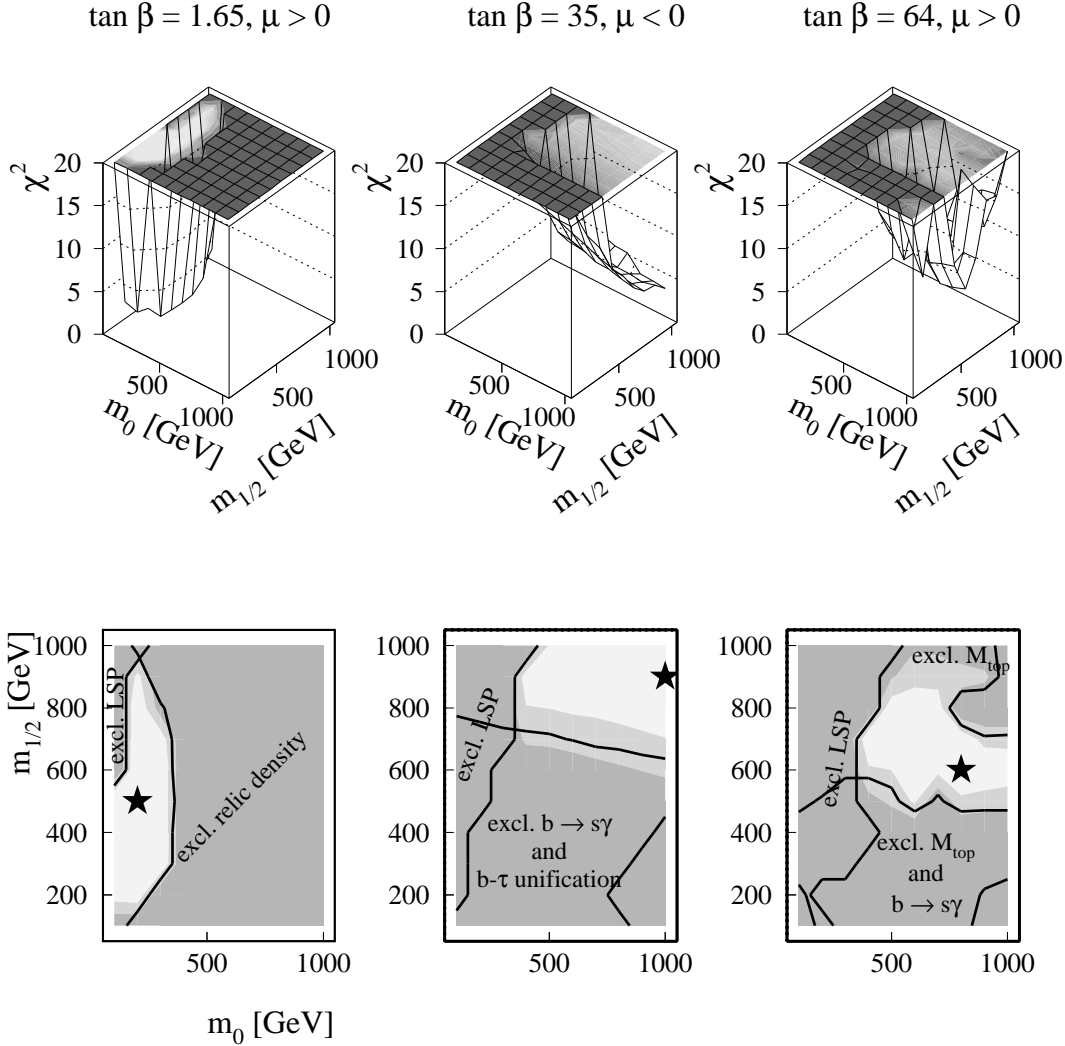,width=0.86\textwidth}
\end{center}
\vspace{-1.cm}
\caption[]{\label{f8}\it The  $\chi^2$-distribution for 
  the low and high $\tb$ solutions.  The different shades
   in the projections  indicate
  steps of $\Delta\chi^2 = 4$, so basically only the light shaded
  region is allowed.
      The stars indicate the optimum solution. 
      Contours enclose domains excluded by the particular
      constraints used in the analysis. From Ref.\cite{wdbpl}.
}
\end{figure}
%
%
  \begin{figure}[t]
   \vspace*{-.8cm}
\begin{center}
    \epsfig{file=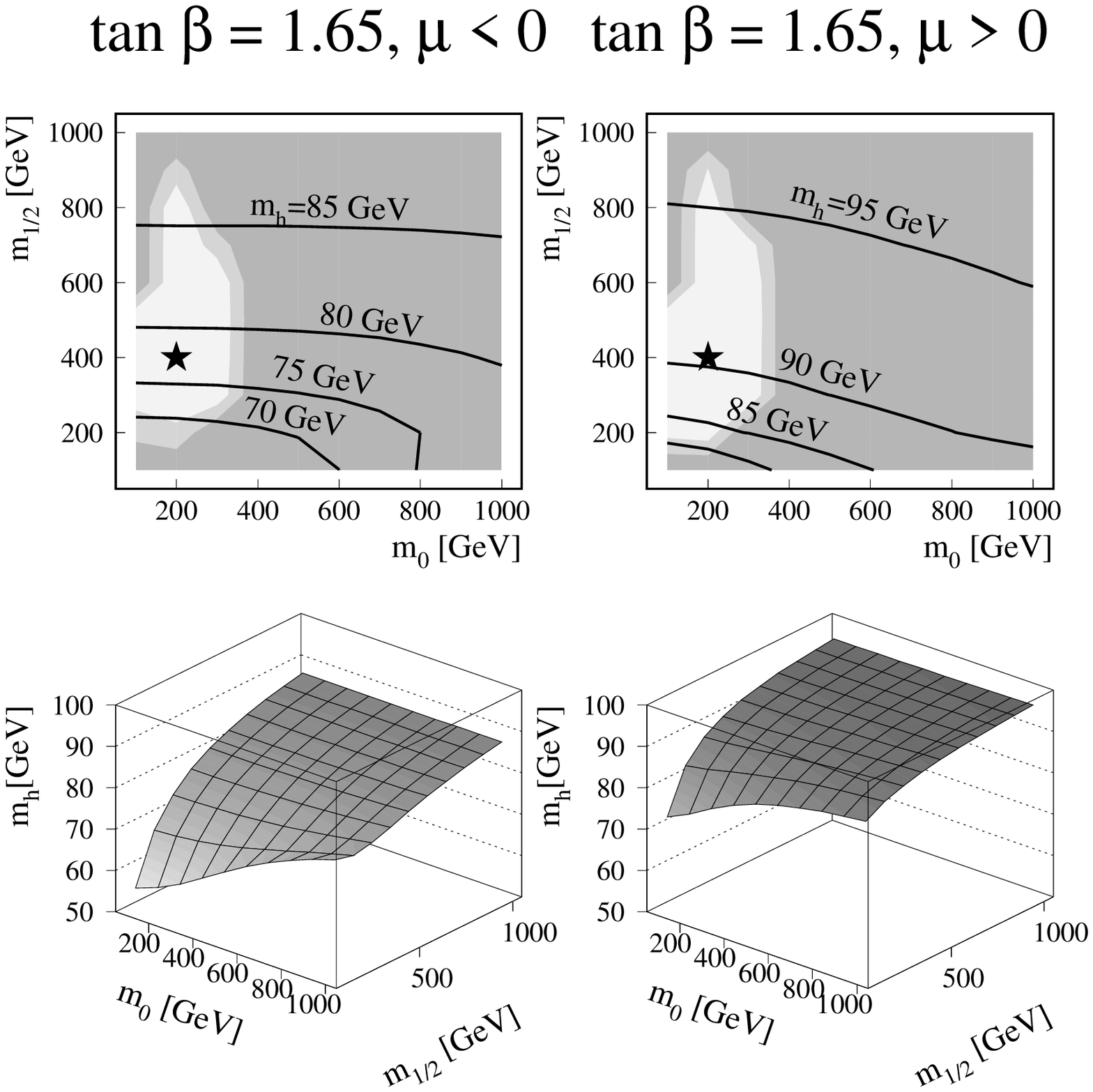,width=0.68\textwidth}
\end{center}
\vspace{-1.0cm}
\caption[]{\label{f10}\it Contours of the  Higgs mass (solid lines) 
  in the $m_0,m_{1/2}$ plane (above)  and the Higgs masses (below) for both 
    signs  of $\mu$  for the low $\tb$ solution $\tb=1.65$ for $m_t=175$ {\rm GeV}.
     The lightly shaded areas correspond to the region allowed by the relic density
     constraint (see Fig. \ref{f8}).
}
\end{figure}
%
%
%

In the SM there is  only one Higgs boson with an arbitrary mass.
Since the decay amplitudes are proportional to the fermion mass,
the branching ratios are predominantly to heavy fermions:
$\approx 84\%$ to $b\overline{b}$ pairs and
$\approx 6\%$ to $\tau\overline{\tau}$ pairs.
Higgs production occurs mainly via Higgs Bremsstrahlung:
$\epem\rightarrow hZ$ with a cross section of the order of a
few tenths of a pb. The
Higgs boson can therefore be searched for in the final states
indicated in Fig. \ref{pie}.
Note that the large fraction of events with b-quarks leads
to very distinct event signatures, which is apparent
from a nice  candidate, shown in Fig. \ref{aleph}.

In the MSSM there are five Higgs bosons: two charged ($\rm H^\pm$),
one heavy neutral ($\rm H$), one light neutral ($\rm h$), and
one neutral pseudoscalar ($\rm A$).
In addition to the SM Higgs bremsstrahlung, one can have 
$\rm hA$ production, if it is kinematically allowed.
This process leads to the final states shown on the right-hand side
of Fig. \ref{pie}.
All these channels have been searched for, but no signal
above the expected background from SM processes has been found.
Consequently, one has to conclude that Higgs bosons are
most likely too heavy to be produced at present centre-of-mass
energies. From a preliminary combination of the 1997 data of the four
LEP experiments at 183 GeV one finds that the SM Higgs mass
has to be above 90.1 GeV\cite{higgswg}.
For the MSSM the lightest Higgs mass,  obtained from
the diagonalization of the mass matrices, is a function
of $\rm \mA$ and $\tb=v_2/v_1$, the ratio of the vacuum
expectation values of the neutral components of the two
Higgs doublets\cite{rev}.
Consequently, the
lower limit on the Higgs mass depends
on $\tb$ and $\rm \mA$. This dependence on three variables is
difficult to depict. Therefore one usually shows the excluded regions
 in the two-dimensional ($ \mh,\tb$) and ($ \mA,\tb$) planes, whilst
 varying the third parameter ($0.5<\tb<50$ and $\rm 0<\mA<2~ TeV$).
 The results are shown in
 Fig. \ref{limits}.
At low $\tb$ the $\rm \mh$ limit  is close to the SM one,
but at high $\tb$ the
limit is approximately 10 GeV lower; the $\mA$ limit for any $\tb$
is about 82 GeV. In special cases (region when $\rm \mh\approx \mA\approx 78~GeV$)
the limits are reduced by a few GeV, as found by special
scans over the SUSY parameters, so that 
  lower limits at 95\% C.L. are\cite{higgswg}:
$$\rm \mh>77 ~GeV; ~\mA>78~GeV.$$
\section{Constraints from low energy data and cosmology}
\label{constraints}
In the gravity inspired scenario, the breaking of supersymmetry 
 occurs via universal gravitational interactions,
which leads to universal masses at the GUT scale.
These common masses at the GUT scale completely determine
the low-energy SUSY spectrum from the known radiative corrections,
which lead to running masses as depicted in Fig. \ref{lsplim}.
 
There are nine free parameters (at the GUT scale) in this minimal
scenario:
the GUT scale $\mgut$ and the unified gauge coupling $\agut$,
the Yukawa couplings of the third generation $Y_t^0,~Y_b^0, Y^0_\tau$
(those of the other generations are small, $b-\tau$ unification
presupposes $Y_b^0=Y^0_\tau$), the common masses for 
spin 0 and spin 1/2 sparticles, called $m_0$ and $m_{1/2}$, respectively,
the Higgs mixing parameter $\mu$,
and the trilinear couplings $A^0$. The superscript denotes the GUT scale value. 
These GUT scale parameters are constrained by the low-energy
data:
the running of the gauge couplings determines   
$\mgut$ and $\agut$, the masses of the third generation
quarks and leptons determine the Yukawa couplings,
$b-\tau$ Yukawa unification
yields the preferred
value for $\tb$ and  electroweak symmetry breaking determines
the absolute value of $\mu$.
The trilinear couplings play a minor role, mainly in
the $\bsg$ rate and the mixing in the stop sector.
 Of course, all parameters are correlated, so they can only be determined in a common fit to the data. Since the mass parameters $\mze, \mha$ are strongly correlated, such a fit was performed for all values
 between 100 GeV and 1 TeV in steps of 100 GeV\cite{wdbpl}.
{\protect\footnotesize
\begin{table}[t]
\vspace*{0.27cm}
\renewcommand{\arraystretch}{1.30}
\renewcommand{\rb}[1]{\raisebox{1.75ex}[-1.75ex]{#1}}
\begin{center}
\begin{tabular}{|c|r|r|r|r|}
\hline
 \multicolumn{4}{|c|}{ Fitted SUSY parameters and masses in GeV}&Lower Limits~         \\
\hline
Symbol & \makebox[2.0cm]{\bf{$\tb=1.65$}} & \makebox[2.5cm]{\bf{ $\tb=35.7$}}&\makebox[2.0cm]{\bf{$\tb=63.8$}}&{ 95\%~C.L.}\\
\hline
\vspace{-0.059cm}
 $m_0$,~   $m_{1/2}$   & 200,~ 500   & 1000,~900   & 800,~600   &   \\
\vspace{-0.059cm}
 $\mu(0)$,~$A(0)$  & 1737, 0.   & -938,~1210  & 1605,~696  &   \\
 \hline
\vspace{-0.059cm}
  $\tilde{\chi}^0_1$,~$\tilde{\chi}^0_2$    &  214,~413
& 397,~722      & 201,~279     &30,~-- \\
\vspace{-0.059cm}
  $ \tilde{\chi}^0_3$,~$ \tilde{\chi}^0_4$   &  1028,~1016
& 834,~791     & 523,~234     &--~,~--\\
\vspace{-0.059cm}
  $\tilde{\chi}^{\pm}_1 $,~$\tilde{\chi}^{\pm}_2$ &  413,~1026
& 721,~834  & 220,~523  &90,~-- \\
\hline
\vspace{-0.059cm}
$\tilde{g}$ & 1155 & 1994   & 1363  &250 \\ \hline \vspace{-0.059cm}
$\tilde{t}_1$,~$\tilde{t}_2$   &  1017,~727  & 1765,~1537
 & 1020,~722    &--,~83\\
$\tilde{b}_1$,~$\tilde{b}_2$   &  953,~1010  & 1734,~1782
 & 991,~1088    & 75,~--\\
%
$\tilde{\tau}_1$,~$\tilde{\tau}_2$   &  279,~403  & 888,~1107
 & 240,~705     & 72,~-- \\
\vspace{-0.059cm}
  $     h,~H $  & 92,~1344   & 130,~1092   & 122,~540   
  & 90,~--\\
\vspace{-0.059cm}
  $ A,~H^\pm $  & 1341,~1344  & 1092,~1096  & 540,~547  
  & 78,~60\\
\hline
\end{tabular} \end{center}
\caption[]{\label{t2}\it Values of the fitted SUSY parameters (upper part)
and corresponding SUSY  masses (lower part)  for low and high
$\tb$ solutions. The experimental 95\% C.L. lower limits on
the sparticle masses in the
last column assume the conservative case of right-handed
sfermions being lighter than the left-handed.
The gluino and lightest neutralino limits assume
 gauge mass unification at the GUT scale, so they follow
in the CMSSM basically from the chargino limit, since the mass ratios
of the chargino and LSP are completely determined by the
running masses (see Fig. \ref{lsplim}).
These mass ratios depend slightly on $\tb$, as can be 
seen from the numerical values in the Table.
The neutral  Higgs limit of 90 {\rm GeV} assumes a large
$\mA$, which is the case in the CMSSM (see the last row).
}
\end{table}
}

Only three values of $\tb$ give an
acceptable $\chi^2$ fit for $m_t=174$ GeV, if the Yukawa couplings are constrained
by $Y_b^0=Y_\tau^0$ (see Fig. \ref{f2}).
 The large $\tb$ solutions have the unique
feature of a possible triple Yukawa unification:  
all three Yukawa couplings are 
driven to an approximate fixed point, as shown on the r.h.s. of Fig. \ref{f2}. These low-energy values of the Yukawa couplings
yield approximately the 
correct masses of the leptons and quarks of the third generation
for $\tb=64$.
The difference between the two solutions at high $\tb$, 
corresponding to  opposite signs of $\mu$, stems from finite loop
corrections to the bottom quark mass
involving squark-gluino and stop-chargino
loops.  These corrections are small for low \tb
solutions, but can become as high as 10--20\% 
for the high \tb values\cite{wezp},
since the dominant  corrections are proportional to $\mu\tan\beta$.
Consequently, they change sign for different signs of $\mu$.

%
%
In Fig.~\ref{f8} the total $\chi^2$ distribution is shown as a
function of $\mze$ and $\mha$ for the three values of $\tb$ determined
above from $b-\tau$ unification. 
The areas at low $m_0$ and high $\mha$ are excluded by the LSP constraint,
since in this case the lightest $\tilde{\tau}$ can become the LSP. If R-parity
is conserved, a charged LSP is not allowed, since the vacuum would be filled
with charged relics from the Big Bang.

The relic density constraint excludes $\mze>350 $ GeV for small $\tb$,
as discussed previously\cite{wezp}. For large
$\tb$ the Higgsino mixture of the LSP allows a fast enough decay via s-channel
$Z^0$ exchange, which means the requirement
$\Omega h^2 \le 1$ is easily fulfilled.
The combined requirements
of  the correct \bsg rate and $b-\tau$ unification 
exclude  a large region of parameter space for large $\tb$, as shown by the contours
in the lower part of that figure\cite{wdbpl}.

One observes  $\chi^2$ minima at $\mze,\mha$ around (200,500),
(1000,900), and (800,600) for the different $\tb$ values,
respectively, as indicated by the stars.

Note that the squarks and gluinos are typically above
1 TeV for the high \tb solutions. Furthermore, the minimal $\chi^2$
values are not excellent for high \tb: for 
$\tb=64$ $\chi^2_{min}=6.1$ from the fitted top mass
 ($\mt=189$ GeV), while for $\tb=35$
  $\chi^2_{min}=4.3$ from \bsg. All other $\chi^2$ contributions are 
negligible.
For   $\tb=1.65$ $\chi^2_{min}=1.7$, basically
from the \Bbsg constraint alone. 

Apart from the heavy spectra for large \tb, 
 one has the problem that
the Born level Higgs masses are
strongly negative, as expected
from the fast running of the soft mass terms of the two Higgs doublets, 
$m_1^2$ and $m_2^2$, which receive negative radiative corrections proportional to 
the Yukawa couplings
(see  the running of $m_2^2$ in Fig. \ref{lsplim}).

For low \tb the present limit on the lightest Higgs mass
severely constrains 
the parameter space in Fig. \ref{f10}, which
shows the  excluded regions in the ($m_0,m_{1/2})$ plane for  
different signs of $\mu$.
As mentioned in the introduction
the SM Higgs limit of 90.1 GeV is also valid for the low $\tb$ 
scenario ($\tb<4$)
of the MSSM. As shown in Fig. \ref{f10}, 
this limit  rules out
the $\mu<0$ solution.
However, this figure assumes $\mt=175 \rm{GeV}$.
 The top mass dependence of the Higgs
mass is slightly steeper than  linear in this range.
Adding about one $\sigma$ to the top mass, 
i.e. $\mt=180$ GeV,
implies that for the contours in Fig. \ref{f10} one should
add 6 GeV to the numbers shown.
Even in this case the $\mu<0$ solution is excluded for a large 
region of parameter
space. Only the small allowed  region with
$\mha>700$ GeV is not yet excluded for $\mt=180$ GeV. Note that in this region
the squarks are well above  1 TeV, and therefore the cancellation
of the  quadratic divergencies in the Higgs masses,
which is only perfect if sparticles and
particles have the same masses,  starts to become worrying.
%
For $\mze=1000,\mha=1000$, which corresponds to  squarks masses
of about 2 TeV\footnote{Explicit
analytical expressions for the sparticle masses as a function of the SUSY parameters
can be found in Ref.\cite{wekzp}.}, one finds
for the upper limit on the Higgs mass in the {\it low} $\tb$ scenario:
$$m_h^{max}=97\pm6~{\rm GeV},$$
 where the error is dominated by the uncertainty from the top mass.
If one requires the squarks to be below 1 TeV, these upper limits are reduced by 4 GeV. 

 This CMSSM number  agrees well with the value from
 Casas et al.: $m_h=97\pm2$ GeV\cite{Haber}.
In both analyses  the Renormalization Group Equations
are used to determine the
trilinear coupling
at low energies and $\mu$ from electroweak symmetry breaking (EWSB),
so the mixing in the stop sector is fixed,
once
the sign of $\mu$ is choosen. Furthermore, in both cases
solutions close to the infrared fixed point are considered,
which are required
in the CMSSM by  EWSB.
The error  on the upper limit quoted above is larger than
the one from Casas et al., as
they did not consider the error on the top mass.

For {\it high} $\tb$ the upper limit on the Higgs mass in the CMSSM is:
$$m_h^{max}=120\pm2~{\rm GeV}.$$
The error from the top mass is small in this case,
as the high $\tb$ fits prefer 
top masses around 190 GeV.

%
%
\section{Summary}
Present search limits are starting to constrain the parameter space
of the Constrained Minimal Supersymmetric Model (CMSSM) considerably.
%
The preferred and allowed region of parameter space is the low $\tb$ region with
a positive Higgs mixing parameter, since
the present Higgs limit of 90.1 GeV   excludes the 
$\mu<0$ solution.
%
The high $\tb$ scenarios
have serious 
finetuning problems, as  all Yukawa couplings are large, 
which causes the Higgs masses at tree level to be
strongly negative (typically --1 TeV),
so the radiative corrections have to be positive and very large to offset this large
negative `starting' value. 
 Furthermore, the solutions with minimal $\chi^2$ require  
squarks above 1 TeV,  which causes an additional finetuning problem
because of the non-cancellation of the quadratic
divergencies to the Higgs masses.

In summary, supersymmetry is still the leading candidate
for physics beyond the SM,
as it provides as a GUT a natural  explanation for:
{
\begin{itemize}
\item the different strengths of the strong and electroweak interactions;
\item  the non-integer electric quark charges;  
\item the different mass scales for quarks and leptons;
\item radiative electroweak symmetry breaking,
  thus linking the heavy top mass and $Z^0$ mass;
\item the large amount of cold dark matter in the universe,
  if R-parity is conserved. 
\end{itemize}
}
In addition, the supersymmetric extension of the SM describes the
low-energy electroweak precision data
as well as the SM.
Do alternative theories exist?
For the individual items, 'yes', but
there is no  known theory, which can explain all these
observations at the same time!

Einstein, when asked what he would think if his
General Theory of Relativity
would not be confirmed by experiment, used to answer:
`The Almighty Lord  missed a most wonderful opportunity'.
 I think the same is true for supersymmetry.

 \section*{Acknowledgements} I wish to express my sincere thanks to
 many colleagues
 from the LEP Working Groups on HIGGS, SUSY and Electroweak Fits for
 their help and fruitfull discussions,
 and A. Sopczak for comments on the manuscript.
 

\end{document}